\documentclass[12pt]{iopart}
\usepackage{amssymb}
\usepackage[mathscr]{eucal}
\usepackage{epic}
\usepackage{eepic}
\usepackage[active]{srcltx}
\usepackage{amsfonts}
\usepackage{bm}

   \newtheorem{theorem}{Theorem}

\newcommand{\nc}{\newcommand}
\nc{\BBS}{{\rm BBS}\ } \nc{\g}{\gamma}  \nc{\lm}{\lambda}
\nc{\la}{\lambda} \nc{\bh}{{\bf h}}  \nc{\av}{\prod_{s\,\in\,\ZN}}
\nc{\cR}{{\cal R}} \nc{\kp}{{\varkappa}} \nc{\om}{\omega}
\nc{\qt}{\tilde{q}} \nc{\tp}{\tilde{p}} \nc{\rt}{\tilde{r}}
\nc{\ty}{\tilde{y}} \nc{\tx}{\tilde{x}} \nc{\tQ}{{\widetilde Q}}
\nc{\trh}{\tilde{\rho}}\nc{\ny}{\nonumber}\nc{\lk}{\left(}
\nc{\rk}{\right)} \nc{\Rb}{\right]} \nc{\Lb}{\left[}
\nc{\rb}{\right\}} \nc{\lb}{\left\{} \nc{\hs}{\hspace*{1cm}}
\nc{\hx}{\hspace*{3mm}} \nc{\hq}{\hspace*{6mm}} \nc{\JStP}{{\it J. Stat.
Phys.}}  \nc{\IJMP}{{\it Intern. J. Mod. Phys.}}

\nc{\al}{\alpha}  \nc{\ZN}{\mathbb{Z}_N}
\nc{\bg}{\boldsymbol{\gamma}} \nc{\bdr}{\boldsymbol{\rho}}
\nc{\bu}{{\bf u}} \nc{\bv}{{\bf v}} \nc{\bV}{{\bf V}}
\nc{\rnk}{r_{n,k}} \nc{\lns}{\la_{n,s}} \nc{\lnl}{\la_{n,l}}
\nc{\FD}{{\cal F}} \nc{\lnk}{\la_{n,k}} \nc{\xnl}{x_{n,l}}
\nc{\Psr}{\Psi_{\bdr_n}} \nc{\ap}{a_{n+1}} \nc{\bp}{b_{n+1}}
\nc{\cp}{c_{n+1}} \nc{\dpp}{d_{n+1}}
\nc{\bep}{\bu_{n+1}^{-1}(\ap-\bp\bv_{n+1})} \nc{\Xop}{\mathbf{X}}
\nc{\Zop}{\mathbf{Z}}  \nc\s{{\gamma}}
\def\r#1{(\ref{#1})}

\nc{\ra}{\rangle} \nc{\BAR}{\begin{array}} \nc{\EAR}{\end{array}}
\nc{\bdm}{\begin{displaymath}} \nc{\edm}{\end{displaymath}}
\nc{\be}{\begin{equation}} \nc{\ee}{\end{equation}}
\nc{\ba}{\begin{array}} \nc{\ea}{\end{array}}
\nc{\bea}{\begin{eqnarray}} \nc{\eea}{\end{eqnarray}}

\nc\sip{\gamma^\prime}\nc\ma{a'}\nc\mb{b'}\nc\mc{c\,'}\nc\md{d\,'}
\nc\pra{a''}\nc\prb{b''}\nc\prc{c\,''}\nc\prd{d\,''} \nc{\hu}{{\bf
u}}\nc{\hh}{{\hat h}}\nc{\bl}{\boldsymbol{\lambda}}\nc{\hv}{{\bf v}}
\nc\si{{\mathrm{s}}}

\begin{document}
 \title{Baxter--Bazhanov--Stroganov model: Separation of Variables and Baxter Equation}
\author{G von Gehlen$^\dag$,~ N Iorgov$^\ddag$,~ S Pakuliak$^{\sharp\flat}$~
 and V Shadura$^\ddag$}
\address{$^\dag$\ Physikalisches Institut der Universit\"at Bonn,
Nussallee 12, D-53115 Bonn, Germany}
\address{$^\ddag$ Bogolyubov Institute for Theoretical Physics, Kiev 03143,
Ukraine}
\address{$^\sharp$\ Bogoliubov Laboratory of Theoretical Physics,
Joint Institute for Nuclear Research, Dubna 141980, Moscow region,
Russia}
\address{$^\flat$\ Institute of Theoretical and Experimental Physics,
Moscow 117259, Russia}
\ead{gehlen@th.physik.uni-bonn.de, iorgov@bitp.kiev.ua, pakuliak@theor.jinr.ru,
shadura@bitp.kiev.ua}
\begin{abstract}
The Baxter-Bazhanov-Stroganov model (also known as the $\tau^{(2)}$ model)
has attracted much interest because it provides a tool for solving the integrable
chiral $\ZN$-Potts model. It can be formulated as a face spin model or via cyclic
$L$-operators. Using  the latter formulation and the
Sklyanin-Kharchev-Lebedev approach, we give the explicit derivation of
the eigenvectors of the component $B_n(\la)$ of the monodromy
matrix for the fully inhomogeneous chain of finite length. For the periodic chain
we obtain the Baxter T-Q-equations via separation of variables.
The functional relations for the transfer matrices of the $\tau^{(2)}$ model
guarantee non-trivial solutions to the Baxter equations.
For the $N=2$ case, which is free fermion point of a generalized Ising model,
the Baxter equations are solved explicitly.
\end{abstract}
\hspace*{2.5cm}{\small \today}\hspace*{12mm}   \submitto{\JPA}
\vspace*{-12mm} 

\section{Introduction}
The aim of this paper is the explicit construction of eigenvectors
of the transfer-matrix for the finite-size inhomogenous periodic
Baxter--Bazhanov--Stroganov model (\BBS model) also known as the
$\tau^{(2)}$-model \cite{B_tau,BaxInv,BS,BBP}. This is a $N$-state spin
lattice model, intimately related to the integrable chiral Potts model.
The connection between the 6-vertex model, the \BBS model
and the chiral Potts model gives the possibility
to formulate a system of functional relations \cite{BS,BBP} for the transfer
matrices of these models. Solving these systems is the basic method for
calculating the eigenvalues of the transfer matrix of the chiral Potts
model \cite{BaxtEV}, and under some analyticity assumptions, to derive the
free energy of this model \cite{BaxtEV} and its order parameter \cite{BaxtOP}.

In general for the \BBS model there is no Bethe pseudovacuum state and so the
algebraic Bethe ansatz cannot be used. Therefore, in order to achieve our goal, we shall use the
formulation of the BBS model in terms of cyclic $L$-operators
first introduced by Korepanov \cite{Kore} and Bazhanov-Stroganov \cite{BS}
and adapt the Sklyanin--Kharchev--Lebedev method of
separation of variables (SoV) \cite{Skly1,KarLeb1,KarLeb2,KhLS}
for solving the BBS eigenvector problem. The fusion equations will
provide the existence of solutions to the Baxter equations.

The paper is organized as follows. After defining the \BBS model as a statistical face spin model,
we give the vertex formulation of the model in terms of a cyclic $L$-operator
and conclude the Introduction explaining the two basic steps involved in the SoV
method. Section 2 deals with the solution of the auxiliary problem,
leaving for Section 3 the lengthy inductive proof of the main
formula. The short Section 4 derives the action of the diagonal component $D$ of the
$L$-operator on the auxiliary eigenstates. Then in Section 5 we come
to the periodic model and in deriving the Baxter T-Q-equations
we show the role of the fusion equations for solving these Baxter
equations. In Section 6 we apply these results to the homogenous $N=2$ case and calculate the
eigenvalues and eigenvectors of the homogenous $N=2$ \BBS
model which is the free fermion point of  a generalized Ising model. Section 7 gives our
conclusions. In an Appendix we show a strong simplification
occurring if the \BBS model is homogenous.

\subsection{The \BBS model}

\begin{figure}[ht]
\begin{center}
\renewcommand{\dashlinestretch}{30}
\unitlength=0.04pt
\begin{picture}(8622,4539)(0,-10)
\drawline(2715,3612)(5415,3612)(5415,912)
    (2715,912)(2715,3612)
\drawline(15,3612)(2715,3612)(2715,4512)
\drawline(5415,4512)(5415,3612)(8115,3612)
\drawline(15,912)(2715,912)(2715,12)
\drawline(5415,12)(5415,912)(8115,912)
\dashline{60.000}(15,3612)(2715,912)(5415,3612)(8115,912)
\dashline{60.000}(15,912)(2715,3612)(5415,912)(8115,3612)
\put(4600,2217){\makebox(0,0)[cc]{$m_{k-1}$}}
\put(7200,2217){\makebox(0,0)[cc]{$m_k$}}
\put(1900,2217){\makebox(0,0)[cc]{$m_{k-2}$}} \put(3100,462)
{\makebox(0,0)[cc]{$\g_{k-1}$}} \put(5685,462)
{\makebox(0,0)[cc]{$\g_k$}}
\put(3100,4017){\makebox(0,0)[cc]{$\g'_{k-1}$}}
\put(5730,4017){\makebox(0,0)[cc]{$\g'_k$}} \put  (15,912)
{\makebox(0,0)[cc]{$\bullet$}}
\put(2715,3612){\makebox(0,0)[cc]{$\bullet$}} \put(2715,912)
{\makebox(0,0)[cc]{$\bullet$}}
\put(5415,3612){\makebox(0,0)[cc]{$\bullet$}} \put(5415,912)
{\makebox(0,0)[cc]{$\bullet$}}
\put(8115,3612){\makebox(0,0)[cc]{$\bullet$}} \put(8115,912)
{\makebox(0,0)[cc]{$\bullet$}} \put  (15,3612)
{\makebox(0,0)[cc]{$\bullet$}}
\put(1365,2262){\makebox(0,0)[cc]{$\bullet$}}
\put(4065,2262){\makebox(0,0)[cc]{$\bullet$}}
\put(6765,2262){\makebox(0,0)[cc]{$\bullet$}}
\end{picture}
\end{center}
\caption{\footnotesize{The triangle with vertices markes by the spin
variables $\g_{k-1}$, $\g'_{k-1}$, $m_{k-1}$ correspond to the
function $F'_{k-1}(\g_{k-1}-\g'_{k-1},m_{k-1})$ in (\ref{bbs_weight});
the triangle $\g_{k}$, $\g'_{k}$, $m_{k-1}$ to
$F''_{k}(\g_{k}-s'_{k},m_{k-1})$. }} \label{tri}\end{figure}
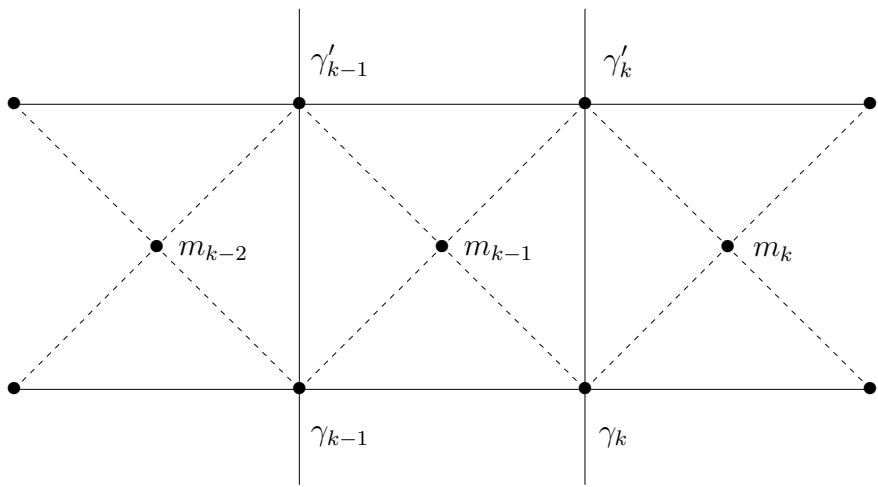

Following the notation of a recent paper of Baxter \cite{B_tau}, we define the BBS model
as a statistical model of short-range interacting spins placed at the vertices of a
rectangular lattice. We label the spin variables
$\si_{x,y}$  by a pair $(x,y)$ of integers: $x=1,\ldots,n,$ and
$y=1,\ldots,m $. Each spin variable $\si_{x,y}$ takes  $N$ values ($N\ge 2$): $0$,
$1,\ldots$, $N-1$. The model shall have $\mathbb{Z}_N$-symmetry and
we may extend the range of the spins $\si_{x,y}$ to all integers identifying
two values if their difference is a multiple of $N$. The model has a chiral restriction on the
values of vertically neighboring spins:
 \begin{equation}
 \label{adj-rule}
  \si_{x,y}-\si_{x,y+1}=0 \mbox{\ or\ } 1 \quad {\rm mod}\ N\ .
\end{equation}
In the following we will consider the spin variables on two adjacent rows:
$(k,\,l)$ and  $(k,\,l+1)$, where $l$ is fixed and $k=1,\ldots,n$.
Let us denote $\si_{k,\,l}=\g_{k}$ and $\si_{k,\,l+1}=\g^\prime_{k}$.
The model depends on the parameters $t_q$ and  $\ma_k\,, \mb_k\,,
\mc_k\,, \md_k$, $\pra_k\,,\prb_k\,, \prc_k\,, \prd_k$,
$k=1,2,\ldots,n$. Each square plaquette of the lattice has the
Boltzmann weight (see Fig.\ref{tri})
\be\label{bbs_weight}
 W_\tau( \g_{k-1},
\g_{k};\sip_{k-1}, \sip_{k}) =  \sum_{m_{k-1}=0}^1
\omega^{m_{k-1}(\sip_{k} - \s_{k-1})} (-\omega t_q)^{ \s_{k} -
\sip_{k}- m_{k-1}}\times \ee
\[
\qquad \qquad \qquad\qquad \times F'_{k-1}( \s_{k-1}- \sip_{k-1},
m_{k-1})
 F''_{k}( \s_{k}- \sip_{k}, m_{k-1}),
 \]
where $\omega=e^{2\pi{\rm i}/N}$, and
\[\fl
F'_{k}(0,0)=1, \qquad F'_{k}(0,1)=-\omega t_q\,\frac{\mc_k}{\mb_k},
\qquad F'_{k}(1,0) = \frac{\md_k}{\mb_k},\qquad F'_{k}(1,1)=-\omega
\frac{\ma_k}{\mb_k},
\]
and expressions for $ F''_{k}( \s_{k}- \sip_{k}, m_{k-1})$ are
obtained from $ F'_k ( \s_{k}- \sip_{k}, m_{k})$ by substitutions:
$\ma_k\,,$ $\mb_k\,,$ $\mc_k\,,$ $\md_k \rightarrow
 \pra_k\,,$ $\prb_k\,, $ $\prc_k\,,$ $ \prd_k\,$.

We will consider the periodic boundary condition: $\s_{n+1}=\s_{1}$,
$\s'_{n+1}=\s'_{1}$, where $n$  is the number of sites on the
lattice along the horizontal axis. The transfer-matrix of the
periodic \BBS model is $N^{n}\times N^{n}$ matrix with matrix elements
\begin{equation}
{\bf t}_{n}({\bg},{\bg}')=\prod_{k=2}^{n+1}W_\tau(
\s_{k-1},
   \s_{k};\sip_{k-1}, \sip_{k}),
 \label{tm}
\end{equation}
labelled by the sets of spin variables ${\bg}=\{\s_1,\s_2,\ldots,\s_n\}$
 and ${\bg}'=\{\s'_1,\s'_2,\ldots,\s'_n\}$ of two neighbour rows.

Considering $m_k$, $k=1,\ldots,n$, in
(\ref{bbs_weight}) as auxiliary spin variables which take the two
values $0$ and $1$, we can rewrite transfer-matrix \r{tm} in a vertex
formulation associating a statistical weight not to the plaquettes but
to vertices each of them relating four spins: $m_{k-1}$, $m_k$,
$\s_{k}$, $\sip_{k}$ (see Fig.~1). Then the weight associated with
the $k$th vertex is
\[
 \ell_k(t_q;m_{k-1}, m_k;\s_{k}, \sip_{k})=
\omega^{m_{k-1}\sip_k-m_{k}\s_k}(-\omega t_q) ^{\s_{k}-
\sip_{k}-m_{k-1}}\times
\]
\be
 \label{tau_2} \,\qquad \qquad \qquad \qquad  \times F''_{k}(
\s_{k}- \sip_{k}, m_{k-1}) F'_{k }( \s_{k }- \sip_{k }, m_{k}).
\ee
and  the transfer-matrix (\ref{tm}) can be rewritten as
\be\label{tm_1}
{\bf t}_{n}({\bg},{\bg}')=\sum_{m_1, ...,\, m_n} \prod_{k=2}^{n+1}
\ell_k(t_q;m_{k-1}, m_k;\s_{k}, \sip_{k}).
\ee

\subsection{The $L$-operator formulation of  \BBS model}

For our construction of the \BBS model eigenvectors we will
use a description of  this model
as a quantum chain model as introduced in \cite{Kore} and \cite{BS}.
To the each site $k$ of the quantum chain we associate
the cyclic $L$-operator acting in a two-dimensional auxiliary space
\\[-6mm]
\be
\label{bazh_strog}L_k(\lm)=\left( \ba{ll}
1+\lm \kp_k \hv_k,\ &  \lm \hu_k^{-1} (a_k-b_k \hv_k)\\ [3mm]
\hu_k (c_k-d_k \hv_k) ,& \lm a_k c_k + \hv_k {b_k d_k}/{\kp_k }
\ea \right)\!\!,\hspace*{4mm} k=1,2,\ldots,n.
\ee
At each site $k$ we define ultra-local Weyl elements $\hu_k$ and $\hv_k$
obeying the commutation rules and normalization
 \be\label{comrel_uv}\fl \bu_j
\bu_k=\bu_k \bu_j\,, \quad\;\bv_j \bv_k=\bv_k
\bv_j\,,\quad\; \bu_j
\bv_k=\om^{\delta_{j,k}}\bv_k\bu_j\,,\quad\; \om=e^{2\pi i/N},
\quad \bu_k^N=\bv_k^N=1.
\ee
In \r{bazh_strog},
$\lm$ is the spectral parameter and we have five
parameters $\:\kp_k,\;a_k,\;b_k,\;c_k,\;d_k\,$ per site. At each site $k$ we
define a $N$-dimensional linear space (quantum space) ${\cal V}_k$
with the basis $ |\g\ra_k$, $\g\in \ZN$ and natural
scalar product $\: _k\langle\g'|\g\ra_k=\delta_{\g',\g}.$ In
${\cal V}_k$ the Weyl elements $\bu_k$ and $\bv_k$ act by the
formulas:
\be\label{uv} \bu_k |\g\ra_k=\om^\g |\g\ra_k\, ,\qquad
\bv_k |\g\ra_k = |\g+1\ra_k\, . \ee

The correspondence between the lattice \BBS model and its quantum chain analog is established
through the relation
\begin{equation}\label{identific}
\ell_k(t_q; m_{k-1}, m_k;\s_{k}, \sip_{k})= \, _k\langle\g| L_k(\lm)_{m_{k-1},m_k}|\g'\ra_k
\end{equation}
and the following connection between the parameters of these models
\[\fl
\lambda=-\omega t_q\,,\quad
\kp_k=\frac{\md_k}{\mb_k} \frac{\prd_k}{\prb_k} \,,\quad
a_k=\frac{\prc_k}{\prb_k}\,,\quad b_k=\omega
      \frac{\pra_k}{\prb_k}\frac{\md_k}{\mb_k}\, ,\quad
c_k=\frac{\prc_k}{\prb_k}\,,\quad
d_k=\frac{\pra_k}{\prb_k}\frac{\md_k}{\mb_k}\,.
\]
We extend the action of the
operators $\bu_k,\;\bv_k\:$ to ${\cal V}^{(n)}={\cal V}_1\otimes{\cal
V}_2\otimes\cdots\otimes{\cal V}_n$ defining this action to be
trivial in all $\:{\cal V}_s\,$ with $\,s\ne k$.
The monodromy matrix for the quantum chain with $n$ sites
is defined as \be \label{mm}
{T}_n(\lm)\,=\,L_1(\lm)\,L_2(\lm)\,\cdots\, L_n(\lm)= \lk \ba{ll}
A_n(\lm)& B_n(\lm)\\ [2mm]
C_n(\lm)& D_n(\lm) \ea \rk. \ee  The transfer-matrix (\ref{tm_1}) is
obtained taking the trace in the auxiliary space
 \be \label{tr_matl} {\bf t}_{n}(\lambda)\:=\:\tr\:
T_n(\lm)\:=\:A_n(\lm)+D_n(\lm)\,.
\ee
This  quantum chain is
integrable because the $L$-operators \r{bazh_strog} are
intertwined by the twisted 6-vertex $R$-matrix at root of unity
\be
{R}(\la,\nu)\;=\;\lk\begin{array}{cccc} \la-\om\nu & 0 & 0 & 0 \\
[0.3mm] 0 & \om(\la-\nu) & \la (1-\om) & 0 \\ [0.3mm] 0 & \nu (1-\om)
& \la-\nu & 0 \\ [0.3mm] 0 & 0 & 0 & \la-\om\nu \end{array}\rk\!,
\ee
\be R(\la,\nu)\,  L^{(1)}_k\, (\la) L^{(2)}_k(\nu)=
L^{(2)}_k(\nu)\, L^{(1)}_k(\la)\, R(\la,\nu), \label{rll} \ee
where
$\:L^{(1)}_k(\la)= L_k(\la)\otimes\mathbb{I}$,
$L^{(2)}_k(\la)=\mathbb{I} \otimes L_k(\la)$. Relation
(\ref{rll}) leads to  $\:[{\bf t}_{n}(\lambda),{\bf t}_{n}(\mu)]=0\;$
and so $\;{\bf t}_n(\la)\,$ is the generating
 function for the commuting set of non-local and
non-hermitian Hamiltonians ${\bf H}_0,\ldots,{\bf H}_n$
of the model:
\be \label{tr_mat_bs} {\bf t}_n(\lm) \:=\:{\bf
H}_0\,+{\bf H}_1\lm\,+\cdots\,+{\bf H}_{n-1}\lm^{n-1}\,+{\bf
H}_{n}\lm^{n}.
\ee
The lowest and highest Hamiltonians can be easily written explicitly in
terms of the global $\ZN$-charge rotation operator $\;\bV_n$
\be \label{h0}\fl {\bf
H}_0\,=\,1\,+\bV_n\prod_{k=1}^n\frac{b_k d_k}{\kp_k};\qquad
{\bf H}_n\,=\,\prod_{k=1}^n\: a_k
c_k\,+\,\bV_n\prod_{k=1}^n \kp_k;\qquad
\bV_n\;=\;\bv_1 \bv_2 \cdots \bv_n. \label{Znc}
\ee
It also follows from the intertwining relation (\ref{rll}) that $B_n(\lm)$
is the generating function for another commuting set of operators
${\bf h}_1,\ldots,{\bf h}_n$: \be\label{Bpol} \left[ B_n(\lambda),
B_n(\mu)\right]=0,\qquad B_n(\lm)= {\bf h}_1 \lm + {\bf h}_2 \lm^2
+\cdots + {\bf h}_{n}\lm^{n}\,  . \ee

The great interest in the \BBS chain model is due to its relation to the
integrable chiral Potts model. In \cite{BS,Kore} it was observed that besides the
intertwining relations \r{rll},
the $L$-operators  \r{bazh_strog} satisfy a second intertwining relation
in the Weyl quantum space:\\[-4mm]
\begin{eqnarray}
\lefteqn{\sum_{\beta_1\beta_2,j}\!
   \mathsf{S}_{\al_1\al_2;\beta_1\beta_2}(p,p',q,q')
   \;L_{i_1j}^{\beta_1\gamma_1}(\la;p,p')\;L_{j\,i_2}^{\beta_2,\gamma_2}
   (\la;q,q')}\ny\\ [-2mm] \hs\hx=\;\sum_{\beta_1\beta_2,j}\!\!\!
    \; L_{i_1j}^{\al_2\beta_2}(\la;q,q')\;L_{j\,i_2}^{\al_1\beta_1}
    (\la;p,p')\;\:\mathsf{S}_{\beta_1,\beta_2;\gamma_1,\gamma_2}(p,p',q,q'),
    \label{SLL}\end{eqnarray}
if the parameters are chosen as
\be \fl \kp_k=\frac{y_{q_k}
y_{q'_k}}{\mu_{q_k} \mu_{q'_k}};\qquad a_k=x_{q_k};\qquad
b_k=\frac{y_{q'_k}}{\mu_{q_k} \mu_{q'_k}};\qquad c_k=\om
\,x_{q'_k};\qquad d_k=\frac{y_{q_k}}{\mu_{q_k} \mu_{q'_k}},
\label{cp_param}\ee
where $x_{q_k},\;y_{q_k},\;\mu_{q_k}$
(analogously for $x_{q'_k}$ etc.) satisfy the chiral Potts model
constraints \be \fl x_{q_k}^N+\,y_{q_k}^N=\mbox{\sf
k}\,(x_{q_k}^Ny_{q_k}^N+1);\quad
  \mbox{\sf k}\,x_{q_k}^N=1-\mbox{\sf k}'\,\mu_{q_k}^{-N};\quad
  \mbox{\sf k}\,y_{q_k}^N=1-\mbox{\sf k}'\,\mu_{q_k}^N;\quad
  \mbox{\sf k}^2+{\mbox{\sf k}'}^2=1. \label{cpcon}\ee
Here $\mbox{\sf k}\,$ and $\,\mbox{\sf k}'\,$ are temperature parameters.
In \r{SLL} we have written the spin matrix elements of the
$L$-operators with the parametrization \r{cp_param} as
$\;L_{i,j}^{\al,\beta}(\la,q,q')\;$ where $i,j=0,1$ are the
components in the auxiliary space and greek indices
$\al,\beta=0,\ldots,N-1$ denote the components in the quantum space
\r{uv}, suppressing the site index $k$. The matrix $\:\mathsf{S}\:$
turns out to be the product of four chiral Potts--Boltzmann weights
\cite{BS} \be \fl\mathsf{S}_{\al_1\al_2,\beta_1\beta_2}(p,p',q,q')
\;=\;W_{pq'}(\al_1-\al_2)
   W_{p'q}(\beta_2-\beta_1)\bar{W}_{pq}(\beta_2-\al_1)\bar{W}_{p'q'}
   (\beta_1-\al_2).\label{BSS}\ee
In the parametrization \r{cp_param} of the \BBS model there are
various functional relations to the chiral Potts model transfer
matrix which have been used to obtain explicit solutions for the
chiral Potts eigenvalues \cite{BBP}. Only further restricting the
parameters to the ``superintegrable chiral Potts'' case:
\be
a_k\;=\;\om^{-1}\,b_k\;=\;c_k\;=\;d_k\;=\;\kp_k\;=\:1 \label{super}
\ee
allows to solve the \BBS model by algebraic Bethe ansatz, see e.g.
\cite{AMCP,TarasovL,BaxSkew}.
 In this form  Baxter in 1989  first obtained the \BBS model as
an ``inverse'' of the superintegrable chiral Potts model, see
Equations (8.13),(8.14) of \cite{BaxInv}.
 In this paper we shall
follow \cite{B_tau} in not using restrictions like \r{cpcon} on
the parameters $\kp_k,$ $a_k,$ $b_k,$ $c_k,$ $d_k$. We only shall exclude
the superintegrable case \r{super}.

It was shown in the paper \cite{Bugrij} that the $N=2$ \BBS model is equivalent
to the generalized Ising model at the free fermion point. The results of our paper
permit to obtain the transfer-matrix eigenvectors for this model.  Recently,
the interesting
paper \cite{Lisovy} appeared, where  these eigenvectors was constructed  using
the Grassmann functional integral.

\subsection{Functional Bethe ansatz and SoV}\label{SoV}
The construction of common eigenvectors of the set commuting
integrals \r{tr_mat_bs} will be solved in two main steps, which
generally can be formulated as follows:

First, for the given quantum integrable chain type model one has to
find an {\it auxiliary} integrable model such that: 1) the
eigenvectors for original model can be expressed as a linear
combination of the eigenvectors for the auxiliary model
 and 2) the coefficients of this decomposition should factorize into
 products of the single variable functions (phenomenon of ``separation of variables").

Second, the auxiliary problem should be chosen in such a way that
the construction of its eigenvectors is a simple iteration process:
eigenvectors for the auxiliary model of  size $n$ have to be
obtained from the eigenvectors for the model of size $n-1$.

An example of realization of the first step in case of the Toda
chain model was proposed in the paper \cite{Gu81}. The auxiliary
model for the periodic Toda chain was the open Toda chain. In the
paper \cite{Sk85} on the Toda example this approach has been
formalized as  ``the functional Bethe ansatz" \cite{Skly1}. A complete
realization of this step for the periodic Toda chain model can be
found in the paper \cite{KarLeb1}.

Regarding  a recurrent procedure for
eigenvectors of the auxiliary problem, probably the first reference
to this possibility may be found in the series of lectures
\cite{Skly2}. The main idea of this approach can be formulated as
follows. Consider an integrable quantum chain model of size $n$. The
monodromy matrix is the product of $n$ $L$-operators. We decompose a
system into two subsystem of the sizes $n_1$ and $n_2$ such that
$n=n_1+n_2$. Suppose we can solve the eigenvalue and eigenvector
problems for the subsystems. In  \cite{Skly2} E.~Sklyanin
claims that there is a relation between the eigenvectors of the
original system and the eigenvectors of its smaller subsystems.
For the open Toda chain model this Sklyanin approach has been realized in the
paper \cite{KarLeb2}.

In our case of the \BBS chain model, the auxiliary model is governed by the
set of commuting integrals \r{Bpol}. So we first solve the problem
of finding the eigenvectors for these integrals. Then we shall show
that  the eigenvectors for the operators \r{tr_mat_bs} can be
constructed  as linear combinations of the eigenvectors of the set
\r{Bpol}. The multi-variable coefficients of this decomposition
admit the separation of variables and can be written as products of
single variable functions, each satisfying a Baxter difference
equation. We shall obtain this Baxter equation for generic $N$ and
solve it explicitly for $N=2$ corresponding to the free fermion
point of the generalized Ising model \cite{Bugrij}.
Note that   the eigenvectors of
the commuting set of operators which come from the generating polynomial
 $A_n(\lambda)$ ($[A_n(\lambda),A_n(\mu)]=0$) were found in paper \cite{Iorgov}.

\section{Eigenvectors of $B_n(\lm)$.}
\subsection{Consequences of the $RTT$ relations}
We start with the second step of the program described in Section \ref{SoV}.
Following \cite{Skly1,Skly2,KarLeb2},
we first construct eigenvectors of  $B_n(\lm)$.
According to (\ref{Bpol}), any common eigenvector of the commuting set of
operators $\,{\bf h}_1,\ldots,{\bf h}_{n}\,$
is eigenvector of $B_n(\lm)$ and the eigenvalue is a polynomial in $\lm$.
Factorizing this polynomial we get\\[-6mm]
\be\label{eig_b}
B_n(\lambda)\,\Psi_{\bl}\: =\:\lm\lm_0\prod_{k=1}^{n-1}\,(\lm-\lm_k)\,\Psi_{\bl};
\qquad {\bl}=(\lambda_0, \lambda_1,\ldots,\lambda_{n-1}),\ee
where we labelled the eigenvector $\Psi_{\bl}$ by the normalizing factor $\lm_0$
and the $n-1$ non-vanishing zeros $\lm_1,\lm_2,\ldots,\lambda_{n-1}$ of
the eigenvalue polynomial.

Now the intertwining relations \r{rll} tell us
that if $\Psi_{\bl}$ is eigenvector of $B_n(\lm)$, then by applying repeatedly
the operators $A(\lm_j)$ and $D(\lm_k)$  $(j,k=1,\ldots,n-1)$ and $\bV_n$ \r{Znc},
we can generate a whole set of $N^n$ eigenvectors of $B_n(\lm)$.

The   intertwining relations (\ref{rll}) give
\bea\fl\label{BA}
(\lambda-\omega\mu)A_n(\lambda)B_n(\mu)&=&\omega(\lambda-\mu)
B_n(\mu)A_n(\lambda)+\mu(1-\omega)A_n(\mu)B_n(\lambda)\\[1mm]
\fl(\lambda-\omega\mu)D_n(\mu)B_n(\lambda)&=&
\omega(\lambda-\mu)B_n(\lambda)D_n(\mu)+\lambda(1-\omega)D_n(\lambda)B_n(\mu)\label{BD}.\eea
Fixing $\;\lm=\lm_k$, $\;k=1,\ldots, n-1\:$ in  (\ref{BA}) and
acting by it on $\Psi_{\bl}$ we obtain
\be
B_n(\mu)\left(A_n(\lambda_k)\Psi_{\bl} \right)=\mu\lambda_0
( \mu-{\omega}^{-1} {\lambda_k}) {\textstyle \prod_{s\neq k}} ( \mu-\lambda_s)\:
\left(A_n(\lm_k)\Psi_{\bl}\right)\,.\ee
This means that up to a constant
\be\label{alm}
A_n(\lambda_k)\Psi_{\bl} \sim \Psi_{\lambda_{\,0}, \ldots ,\,
 {\omega}^{-1} \lambda_k , \ldots ,\, \lambda_{n-1} }\,.
\ee
Similarly, from \r{BD} we can get
\be\label{dlm}
D_n(\lambda_k)\Psi_{\bl} \sim \Psi_{ {\omega}^{-1}{\lambda_{\,0}} , \ldots ,
 \,{\omega}{ \lambda_k} , \ldots , \,\lambda_{n-1} }.
\ee
(The proportional factors will be obtained later in \r{Almk} and \r{Dlmk}).
Furthermore, acting by (\ref{BA}) on $\,\Psi_{\bl}$ and extracting
coefficients of $\:\lm^{n+1}\mu^n\,$ we have
\be\label{vvv}
\bV_n \Psi_{\bl}\sim \Psi_{ {\omega}^{-1}{\lambda_{\,0}} , \ldots ,
 \,{ \lambda_k} , \ldots , \,\lambda_{n-1} }\,.
\ee

We see that the operators $A_n(\lm)$ and $D_n(\lm)$
at the eigenvalue zeros $\lm_k$ of $B_n(\lm)$, together with the charge rotation operator
$\bV_n=\bv_1 \bv_2 \cdots \bv_n$ act as cyclic ladder operators on the eigenvectors
of $B_n(\lm)$.
So the eigenvalues of $B_n(\lm)$ can be written
\be\label{iBlm}
B_n(\lm)\Psi_{\bdr_n} =\lm\, r_{n,0}\,\om^{-\rho_{n,0}}
\prod_{k=1}^{n-1}\lk\lm+r_{n,k}\om^{-\rho_{n,k}}\rk\Psi_{\bdr_n}\,,
\ee
where $r_{n,s}$, $s=0,\ldots,n-1$ is a set of constants
and we shall use the phases
\be    \bdr_n=(\rho_{n,0},\ldots,\rho_{n,n-1})\in(\ZN)^n\label{br}\ee
as new labels of the eigenvectors.
The fact that the eigenvectors of the operator
$B_n(\lambda)$ can be considered as dependent only on the integer phases
of the roots
\be\label{roots}
\lambda_{n,k}=-r_{n,k}\om^{-\rho_{n,k}}
\ee is a common property
of the root of unity integrable models. The amplitudes $r_{n,k}$ of these roots are
fixed by some``classical" procedure
which will be described below. In some cases this procedure becomes
an classical integrable system naturally incorporated into the quantum system (see
\cite{GPS} and references therein).

\subsection{One- and two-site eigenvectors for the auxiliary problem}

We now start to solve the auxiliary problem, which is
to compute the eigenvectors of $B_n(\lm)$ in the basis ${\cal V}^{(n)}$.
We shall adapt the recursive procedure of Kharchev and
Lebedev \cite{KarLeb2} to the \BBS chain model.

In our root of unity case a very important role will be played by the
cyclic function $\:w_p(\g)\:$ \cite{BB} which
depends on a $\ZN$-variable $\g$ and on a point $p=(x,y)$
restricted to the Fermat curve $x^N+y^N=1$. We define $w_p(\g)$ by the difference equation
\be  \fl
\frac{w_p(\g)}{w_p(\g-1)}\:=\:\frac{y}{1\,-\,\om^\g\,x}\,;\qquad
x^N\,+\,y^N\,=\,1\,;\qquad \g\in\ZN;\hq\quad w_p(0)=1\,. \label{Fermat} \ee
The Fermat curve restriction guarantees the cyclic property $\;w_{p}(\g+N)=w_{p}(\g).$
The function $\:w_p(\g)\:$ is a root of unity analog of the $q$-gamma function.

It is convenient to change the bases in the spaces ${\cal V}_k$.
Instead of $|\g\ra_k$, $\g\in \ZN$, we will use the vectors
\be\label{psik}
\psi^{(k)}_{\rho}=
\sum_{\g\in \ZN} w_{p_k}(\g-\rho)|\g\ra_k,\qquad \rho\in \ZN\,,
\ee
which are eigenvectors of the upper off-diagonal matrix element
$\:\lm\bu_k^{-1} (a_k-b_k \bv_k)\,$ of the operator $L_k$:
\bea \fl \la\,\bu_k^{-1} (a_k-b_k \bv_k)\, \psi^{(k)}_{\rho}\ny\\
 \hspace*{-1cm}=\: \la\,a_k\!\!\sum_{\g\in
\ZN} w_{p_k}(\g-\rho)\om^{-\g}|\g\ra_k-\la\,b_k\!\!\sum_{\g\in
\ZN} w_{p_k}(\g-\rho)\om^{-(\g+1)}|\g+1\ra_k
\ny\\
\hspace*{-1cm}=\: \la\,a_k\!\!\sum_{\g\in
\ZN} w_{p_k}(\g-\rho)\om^{-\g}|\g\ra_k-\la\,b_k\!\!\sum_{\g\in
\ZN} w_{p_k}(\g-\rho-1)\om^{-\g}|\g\ra_k
\ny\\
 \hspace*{-1cm} =\:\la\,\sum_{\g\in \ZN} w_{p_k}(\g-\rho)\left[
\lk a_k -\frac{b_k}{y_k}\rk\om^{-\g}+b_k\frac{x_k}{y_k}\om^{-\rho}\right]|\g\ra_k
=\la\,r_k\,\om^{-\rho}\,\psi_\rho^{(k)}.
  \label{efb} \eea
In the first step we used \r{uv} and to obtain the last line we used \r{Fermat} with
 $y_k=b_k/a_k$, $r_k=x_k a_k$. The Fermat curve restriction for $p_k=(x_k,y_k)$ gives
 $r_k^N=a_k^N-b_k^N$. We see that if $r_k=0$ (in particular, in the superintegrable
 case \r{super}) it leads to
$x_k=0,\;y_k=1$. In this case \r{psik} does not give a new basis in ${\cal V}_k$.
This is a reason why we exclude values of the parameters which lead to
the degeneration of the cyclic  function $w_p(\gamma)$.

 This sequence of operations applied in \r{efb} will be performed rather often in the
 following derivations. The application of $\bv_k$ shifts the
 spin index. This is compensated by the shift of the summation variable, which
 results in an opposite shift of the argument of $w_p$. This in turn is
 removed using \r{Fermat}.

The operator $\bv_k$ shifts the index of $\:\psi_\rho^{(k)}$:
\be \bv_k\psi_\rho^{(k)}=\sum_{\g\in
\ZN} w_{p_k}(\g-\rho)|\g+1\ra_k=
\sum_{\g\in\ZN} w_{p_k}(\g-\rho-1)|\g\ra_k= \psi_{\rho+1}^{(k)}\,.
\label{vpsi}\ee
Using \r{efb} for $k=1$ and comparing to (\ref{iBlm}), we write one-site
eigenvector as $\Psi_{\rho_{1,0}}:= \psi^{(1)}_{\rho_{1,0}}$. With $\;r_{1,0}\,=r_1\:$
we have  \be\label{AB1}
B_1(\lm)\Psi_{\rho_{1,0}}=\lm\: r_{1,0}\: \om^{-\rho_{1,0}}\, \Psi_{\rho_{1,0}}\,,
\quad
A_1(\lm)\Psi_{\rho_{1,0}}=\Psi_{\rho_{1,0}}+\lm\kp_1 \Psi_{\rho_{1,0}+1}\,.
\ee

The construction of the two-site eigenvectors $\,\Psi_{\bdr_2}\,$ will show us
the first step of the recursive method.
In accordance with (\ref{iBlm}) we are looking for
eigenvectors $\Psi_{\bdr_2}$,
$\:(\bdr_2\equiv(\rho_{2,0},\rho_{2,1})\in{\mathbb Z}_N\times\ZN)$
of the two-site operator $B_2(\la)$, which should satisfy
\be\label{B2}
B_2(\la)\Psi_{\bdr_2}=\la\,r_{2,0}
        \om^{-\rho_{2,0}}(\la+r_{2,1}\om^{-\rho_{2,1}})\Psi_{\bdr_2}\,.
\ee
We suppose that $\Psi_{\bdr_2}$ can be
written as a linear combinations of products of one-site eigenvectors
\be \label{Psi2}
\Psi_{\bdr_2}\;=\;
\sum_{\rho_1,\,\rho_2\in \ZN}Q(\rho_1,\rho_2| \bdr_2)\:
\psi^{(1)}_{\rho_1}\!\otimes \psi_{\rho_2}^{(2)}\,.
\ee
Using (\ref{AB1}) and (\ref{efb}), the matrix $Q$ can be calculated as follows:
\bea
\fl B_2(\la)\Psi_{\bdr_2}= \left(A_1(\lm)\, \lm\,\bu_2^{-1} (a_2-b_2 \bv_2)+
B_1(\lm) (\lm a_2 c_2+b_2\,d_2\bv_2/\kp_2)\right) \Psi_{\bdr_2}=\ny\\ [2mm]
\hspace*{-1.5cm}   =
\sum_{\rho_1, \,\rho_2 } \lb Q(\rho_1,\rho_2|\bdr_2 ) \lk \la r_2\om^{-\rho_2}
+\la^2 a_2c_2 r_1\om^{-\rho_1}\rk +Q(\rho_1-1,\rho_2|\bdr_2 )\la^2
\kp_1 r_2\om^{-\rho_2}\right.
\ny \\ [-2mm]
 \hspace*{1.5cm}  +  \left.Q(\rho_1,\rho_2-1|\bdr_2 )
\frac{b_2d_2}{\kp_2}\la r_1\om^{-\rho_1}\!\rb
\psi^{(1)}_{\rho_1}\!\otimes \psi^{(2)}_{\rho_2}.\label{B2calc}
\eea
Comparing powers of the spectral parameter $\lm$ in (\ref{B2calc}) and
in (\ref{B2}), together with (\ref{Psi2}), we get
\be
\hspace*{-1.5cm}\lk r_{2,0}\,\om^{-\rho_{2,0}}-
a_2\,c_2\,r_1\,\om^{-\rho_1}\rk Q( \rho_1,\rho_2|\bdr_2) \;=
\;  \kp_1 r_2\om^{-\rho_2} Q( \rho_1-1,\rho_2|\bdr_2),\label{twop1}
\ee
\be
\hspace*{-1.5cm}\lk r_{2,0}r_{2,1}\om^{-\rho_{2,0}-\rho_{2,1}}
-r_2\om^{-\rho_2}\rk Q( \rho_1,\rho_2|\bdr_2 )=
\frac{b_2d_2}{\kp_2}r_1\om^{-\rho_1}Q(\rho_1,\rho_2-1|\bdr_2 ).\label{diff2}
\ee
The difference equations (\ref{twop1}) and (\ref{diff2}) have the solution
\be \label{twop3}
  Q(\rho_1,\rho_2|\bdr_2)\!=\!\frac{\om^{-(\rho_{2,0}+\rho_{2,1}-\rho_1)
 (\rho_{2,0} -\rho_2)}}
{w_{p_{\,2,\,0}}(\rho_{2,0}-\rho_1-1) w_{\tilde p_{\,2}}(\rho_{2,0}+\rho_{2,1}-\rho_2-1 )},
\ee
where $p_{2,\,0}=(x_{2,\,0},y_{2,\,0})$,
$ \tilde p_{2}=( \tilde x_{2},\tilde y_{2})$ and
 \be
 \hspace*{-1.5cm} x_{2,\,0}=a_2\,c_2\frac {r_1}{r_{2,\,0}}\,, \hq
 y_{2,\,0}=\kp_1\,\frac {r_2}{r_{2,\,0}}\,,\hq
 \tilde x_{2}= \frac {r_2}{r_{2,\,0}\, r_{2,\,1}}\,, \hq
\tilde y_{2}= \frac{b_2\,d_2}{\kp_2}\frac {r_1}{r_{2,\,0}\, r_{2,\,1}}\,.\label{ptwo}
 \ee
The parameters $r_{2,0}$ and $r_{2,1}$ are determined from the
condition that the points $p_{2,0}$ and $\tilde p_{2}$ belong to the Fermat curve.

One can proceed this way to construct $n$-site eigenvectors of the auxiliary
problem. In fact, in order to see the general structure emerging, one has to go to
the four-site case. We shall not do this here, but rather in Section~\ref{eigenvectors}
we shall prove the general result by induction. This proof will use recursive relations
between amplitudes $r_{n,k}$, $k=0,1,\ldots,n-1$ formulated in the
Subsection~\ref{subsec24}. Fermat parameters $p=(x,y)$ of
the cyclic functions $w_{p}(\rho)$ appearing in our construction will
depend on these amplitudes. Compatibility conditions between recursive relations
for the amplitudes and Fermat curve equation $x^N+y^N=1$ can be formulated
as a ``classical'' BBS chain model \cite{GPS}.
This model will be formulated in the next Subsection
using an averaging procedure for the cyclic $L$-operators \r{bazh_strog} given in
\cite{Tarasov}.

\subsection{Determination of the parameters  $r_{m,s}$}\label{recinh}

Let us define the ``classical'' counterpart
$\,{\cal O}(\lm^N)\,$ of a quantum cyclic operator $\,O(\lm)\,$ using
averaging procedure \cite{Tarasov}
\be {\cal O}(\lm^N)\;=\;\langle\,O\,\rangle(\lm^N)=\;{\textstyle \av} O(\om^s\lm)
\label{ave}\ee
and apply this procedure to the entries of the quantum $L$-operator \r{bazh_strog}.
Denote the result by $\mathcal{L}_k(\lm^N)$
\be\label{Lclass}
\fl \mathcal{L}_k(\lm^N)\;=\;\left(\ba{cc} \langle\, L_{00}\,\rangle&
         \langle\,L_{01}\,\rangle \\ [2mm]
         \langle\,L_{10}\,\rangle & \langle\,L_{11}\,\rangle\ea\right)
=\left( \ba{cc}1-\epsilon \kp_k^N \lm^N &\;\; -\epsilon\lm^N(a_k^N-b_k^N)\\ [2mm]
            c_k^N-d_k^N &\;\; b_k^N d_k^N/\kp_k^N-\epsilon \lm^N a_k^N c_k^N\ea \right)\ee
where $\epsilon=(-1)^N$, and call it as the ``classical'' $\mathcal{L}$-operator
of the classical BBS model. Accordingly, the classical
monodromy ${\cal T}_m$ for the $m$-site chain is
\be \label{Tclass}
\mathcal{T}_m(\lm^N)\:=\;{\cal L}_1(\lm^N)\: {\cal L}_2(\lm^N) \cdots\:
{\cal L}_m(\lm^N)\:=\:\left(\ba{cc}
{\cal A}_m(\lm^N) &  {\cal B}_m(\lm^N)\\ [1mm]
{\cal C}_m(\lm^N) &  {\cal D}_m(\lm^N)
\ea\right)\ee
where the entries are polynomials of $\lm^N$.
 By Proposition~1.5 from \cite{Tarasov}, these
polynomials coincide with averages $\langle{A}_m\rangle$, $\langle{B}_m\rangle$,
$\langle{C}_m\rangle$ and $\langle{D}_m\rangle$ of the entries of \r{mm}.
This proposition provides a tool for finding the $N$-th powers of the amplitudes $r_{m,s}$:
applying \r{ave} to \r{iBlm} we obtain
\be
 \lefteqn{{\cal B}_m(\lm^N)=(-\epsilon)^m \lm^N r_{m,0}^N \prod_{s=1}^{m-1}(\lm^N-
\epsilon\, r_{m,s}^N)\,.}\label{Brec}
\ee
This relation together with \r{Lclass} and \r{Tclass} allows to find $r_{m,s}^N$ in terms of
the parameters $a_k^N$, $b_k^N$, $c_k^N$, $d_k^N$ and $\kp_k^N$, $k=1,\ldots,m$.
The problem of finding the amplitudes $r_{m,s}$ is reduced
to the problem of solving a $(m-1)$-th degree algebraic relation.
As shown in the Appendix, in
the case of the homogeneous \BBS chain model
the problem is reduced to solving a quadratic equation only.
The described procedure gives the amplitudes $r_{m,s}$ up to some roots of unity.
In fact we can fix these phases arbitrarily because this leads just to relabeling of the
eigenvectors. In what follows we suppose that we fixed some solution
$\{r_{m,s}\}$ in terms of the parameters $a_k^N$, $b_k^N$, $c_k^N$, $d_k^N$ and $\kp_k^N$.

Let us give a recursive description for ${\cal B}_m(\lm^N)$.
From \r{Tclass}, we immediately read off the recursion relations
\[\fl
{\cal A}_m(\lm^N)= (1-\epsilon \kp_m^N \lm^N ) {\cal A}_{m-1}(\lm^N)
+(c_m^N-d_m^N) {\cal B}_{m-1}(\lm^N)\,,
\]
\be\label{BAB}\fl
{\cal B}_m(\lm^N)= -\epsilon\lm^N(a_m^N-b_m^N) {\cal A}_{m-1}(\lm^N)
+(b_m^N d_m^N/\kp_m^N-\epsilon \lm^N a_m^N c_m^N) {\cal B}_{m-1}(\lm^N)\,.
\ee
Combining these two relations we get
\be\label{calAB}
{\cal A}_m(\lm^N)\,=\,
\frac{\epsilon\,\kp_m^N\lm^N-1}{\epsilon\,\lm^N\,(a_m^N\,-b_m^N)}\: {\cal B}_m(\lm^N)
\,+\,\frac{\det\, {\cal L}_m(\lm^N)}{\epsilon\,\lm^N\,(a_m^N\,-b_m^N)}\: {\cal B}_{m-1}(\lm^N),
\ee
where \\ [-8mm] \be \det \:{\cal L}_m(\lm^N)\;=\;(d_m^N\,-\epsilon\,\lm^N\,c_m^N\,\kp_m^N)\,
(b_m^N\,-\epsilon\,\lm^N\,a_m^N\,\kp_m^N)/\kp_m^N. \label{detlN}\ee
Substituting ${\cal A}_{m-1}$ from this equation with $m$ replaced by $m-1$ into
\r{BAB}, we
obtain a three-term recursion for $\:{\cal B}_m(\lm^N)$:
\bea \fl {\cal B}_m(\lm^N)&=&\lk\frac{r_m^N}{r_{m-1}^N}(1-\epsilon\,\kp_{m-1}^N\lm^N)+
\,b_m^N\, d_m^N/\kp_m^N\,-\epsilon\, \lm^N a_m^N\, c_m^N\rk {\cal
B}_{m-1}(\lm^N)+ \ny\\
\fl &&\hspace*{-20mm}+\:\frac{r_m^N}{r_{m-1}^N\kp_{m-1}^N}(b_{m-1}^N -\epsilon
\lm^N a_{m-1}^N \kp_{m-1}^N ) (\epsilon \lm^N c_{m-1}^N
\kp_{m-1}^N - d_{m-1}^N)\: {\cal B}_{m-2}(\lm^N), \quad m\ge 2,
\label{relBm} \eea
where we abbreviated
\be r_m^N\;=\;a_m^N\,-\,b_m^N. \label{rmN}\ee
To define $\:{\cal B}_m(\lm^N)$ by \r{relBm} we have to provide the initial values
\[
{\cal B}_1(\lm^N)=-\epsilon\, \lm^N r_1^N; \hq\qquad
{\cal B}_0(\lm^N)=0.
\]

\subsection{Fermat curve points appearing in the construction of
                      the eigenvectors of $B_n(\lm)$}
                      \label{subsec24}

As we have seen in the case of the two-site chain,
the formulas \r{Psi2}, \r{twop3} for the eigenvectors are given in terms of the points
$p_{2,\,0}$ and $ \tilde p_{2}$ on the Fermat curve. The coordinates of these points are
fixed by the values of amplitudes  $r_{2,0}$, $r_{2,1}$, see \r{ptwo}.
In the $n$-site case  four types of such points will appear:
\be  \fl \tilde p_m=(\tilde
x_m,\tilde y_m);\quad
p_{m,s}=(x_{m,s}, y_{m,s});\quad\tilde p_{m,s}=(\tilde
x_{m,s},\tilde y_{m,s});\quad
p^{m,s}_{m',s'}=(x^{m,s}_{m',s'},y^{m,s}_{m',s'}).\label{pxy}\ee
The coordinates of these points are expressed in the terms of amplitudes
$r_{m,s}$, $m=1,\ldots,$ $n$, $s=0,\ldots,$ $m-1$ (defined as {\it some} solutions of
equations \r{Brec}, $m=1,\ldots,$ $n$) by
\be \fl x^{m,s}_{m',s'}=r_{m,s}/r_{m',s'},\quad x_{m,s}=a_m \kp_m
r_{m,s}/b_m,\quad \tilde x_{m,s}=d_m/(\kp_m c_m r_{m,s}),\quad s,s'\ge 1.
\label{xxx}\ee
The corresponding $y^{m,s}_{m',s'}$, $\tilde y_{m,s}$ are defined
by the only condition on $p^{m,s}_{m',s'}$, $\tilde p_{m,s}$ to belong to the Fermat curve.
The coordinates $y_{m-1,l}$, $1\le l\le m-2$ are defined by
\[\fl
\frac{\tilde r_{m-1} r_{m,0}\,r_{m-1}}{\tilde r_{m-2}\,r_{m-1,0} r_{m}\,
b_{m-1}\,c_{m-1}\, y_{m-1,l}\,\tilde y_{m-1,l}}\;
\]
\be \label{rel_other}
\qquad\qquad \times \prod_{s\ne l}^{m-2} \frac{y^{m-1,l}_{m-1,s}}{y^{m-1,s}_{m-1,l}}\;
\;\frac{\prod_{k=1}^{m-1} \;y^{m,k}_{m-1,l}}{\prod_{s=1}^{m-3}\; y_{m-2,s}^{m-1,l}}\:=\:1\,,
\qquad l=1,\ldots,m-2,
\ee
where
\be \rt_m\,=\,r_{m,0}\,r_{m,1}\,\ldots\, r_{m,m-1}\,.\label{rtilde}\ee
The coordinates of the points $p_{m,0}$ and ${\tilde p}_m$  are defined by
\be x_{m,0} r_{m,0}=r_{m-1,0} a_m c_m,\qquad y_{m,0}
r_{m,0}=\kp_1\kp_2\cdots\kp_{m-1} r_m\ ,\label{xynull}\ee
\be \tilde x_m \tilde r_m =r_m,\qquad \tilde y_m \tilde r_m =b_m
d_m \tilde r_{m-1}/\kp_m\,.\label{xyt}\ee
Formulas \r{rel_other}--\r{xyt}  are result from the solution of the
eigenvector problem \r{iBlm}, see Section~3.

The condition on the points $p_{m-1,l}$ ($1\le l\le m-2$), $p_{m,0}$ and ${\tilde p}_m$ defined
by \r{rel_other}, \r{xynull}, \r{xyt} to belong to the Fermat curve gives
\bea \fl \lefteqn{r_{m-1}^N  \kp_{m-1}^N r_{m,0}^N
 \prod_{k=1}^{m-1} (r_{m-1,l}^N-r_{m,k}^N)=
r_{m}^N r_{m-2,0}^N\:  (b_{m-1}^N -a_{m-1}^N \kp_{m-1}^N r_{m-1,l}^N)\,\times}
\ny\\ [-3mm]\fl &&\label{relotherN}\hs \times\, (r_{m-1,l}^N c_{m-1}^N  \kp_{m-1}^N-
d_{m-1}^N) \prod_{s=1}^{m-3} (r_{m-1,l}^N-r_{m-2,s}^N), \hq
l=1,2,\ldots,m-2, \eea
\be\label{rel0N}
r_{m,0}^N= r_{m-1,0}^N a_m^N c_m^N+\kp_1^N\kp_2^N\cdots\kp_{m-1}^N r_m^N,
\ee
\be\label{reltN}
\rt_m^N\:\equiv\:r_{m,0}^N r_{m,1}^N \cdots r_{m,m-1}^N\:
             =\:r_m^N+b_m^N d_m^N \rt_{m-1}^N/\kp_m^N\,.\ee
In order to relate these relations to the recurrent formulas of the classical BBS model
\r{relBm} we observe that
the relations \r{rel0N} (resp. \r{reltN}) follow from the relations obtained by
the consideration of the highest (resp. lowest) terms in $\lm$ in \r{relBm}
starting from $m=2$. Then, fixing in \r{relBm} $\lm^N$ successively at the $m-2$
non-vanishing zeros of ${\cal B}_{m-1}$,
i.e. putting $\lm^N\,=\,\epsilon r_{m-1,l}^N$,  $l=1,\ldots,m-2,$
we obtain \r{relotherN}. Thus the points $p_{m-1,l}$, $p_{m,0}$ and ${\tilde p}_m$
defined by \r{rel_other}, \r{xynull}, \r{xyt} belong to the Fermat curve automatically.

At the end of this section we would like to mention that the amplitudes $r_{m,s}$ can be found directly
(i.e. not using the results from the previous subsection) from the relations
\r{rel_other}, \r{xynull}, \r{xyt} considered as equations with respect to
$r_{m,s}$ and the coordinates of the Fermat curve points \r{pxy}.
These equations can be solved recursively starting from $m=2$ and taking
$N$-th powers of these relations (see \r{relotherN}, \r{rel0N}, \r{reltN}).
Then the explicit formulas for the eigenvectors from the next section allow
to obtain the Tarasov Proposition~1.5 in \cite{Tarasov}  as a corollary.

\section{Inductive proof of the general solution of the auxiliary problem}
\label{eigenvectors}

Recall from (\ref{AB1}) that the vector
$\Psi_{\rho_{1,0}}:= \psi^{(1)}_{\rho_{1,0}}\in {\cal V}_1$ is
eigenvector for $B_1(\lm)$:
\[B_1(\lm) \Psi_{\rho_{1,0}} = \lm\: r_{1,0}\: \om^{-\rho_{1,0}}\, \Psi_{\rho_{1,0}},\]
and
recall from \r{iBlm}, \r{br} that the eigenvectors $\Psi_{\bdr_n}$ of $B_n(\la)$
were labeled by the vector $\bdr_n=(\rho_{n,0},\ldots,\rho_{n,n-1})\in(\ZN)^n$.
Let us further define:
\be  \trh_n={\textstyle \sum_{k=0}^{n-1}}\;\;\rho_{n,k};\qquad\quad \bdr_n'=
(\rho_{n,1},\ldots,\rho_{n,n-1})\in(\ZN)^{n-1}. \ee
$\bdr_n^{\pm k}\;$ denotes the vector $\;\bdr_n\;$ in which $\rho_{n,k}$ is
replaced by $\rho_{n,k}\pm 1$, i.e.\\[1mm]
$\hs\hs\hx\;\:\bdr_n^{\pm k}=(\rho_{n,0},\ldots,\rho_{n,k}\pm 1,\ldots,\rho_{n,n-1}),\hq
k=0,1,\ldots,n$.\\[-3mm]

The following Theorem~\ref{th1} gives a procedure to obtain the eigenvectors
$\Psi_{\bdr_n}\:\in\:{\cal V}^{(n)}$, $n\ge 2$,
of $\;B_n(\lm)\;$ from eigenvectors
$\Psi_{\bdr_{n-1}}\in\,{\cal V}^{(n-1)}\;$ of $\:B_{n-1}(\lm)\,$
and single site vectors $\psi^{(n)}_{\rho_n}\:\in\:{\cal V}_{n}$
defined by (\ref{psik}). We start from $\Psi_{\rho_{1,0}}$.
As result of the first step of the induction we obtain the two-site
result \r{Psi2} with \r{twop3},\r{ptwo}.

The following theorem is valid provided  $r_m^N\ne 0$,
the polynomials ${\cal B}_m(\lm^N)/\lm^N$, $m=2,\ldots,n$,
have nonzero simple zeros and $\det {\cal T}_n(\epsilon r_{m,s}^N)\ne 0$
(cf. the definition of the $B$-representation in \cite{Tarasov}).

\begin{theorem}\label{th1}
The vector
\be
\Psi_{\bdr_n}=\sum_{\bdr_{n-1}\in (\ZN)^{n-1}\atop \rho_n \in \ZN}
 Q(\bdr_{n-1},\rho_n|\bdr_n) \Psi_{\bdr_{n-1}}\otimes
\psi^{(n)}_{\rho_n} \label{PSI}\ee
where \\ [-13mm]
\bea
Q(\bdr_{n-1},\rho_n|\bdr_{n})&=&\frac{\om^{(\tilde \rho_n-
\tilde \rho_{n-1}) (\rho_n-\rho_{n,0})}}
{w_{p_{n0}}(\rho_{n,0}-\rho_{n-1,0}-1) w_{\tilde p_n}(\tilde
\rho_n- \rho_n-1)}\: \times \ny \\ [2mm]  \label{QQQ} &&\hspace*{-2cm}\times\:
\frac{ \prod_{l=1}^{n-2} \prod_{k=1}^{n-1}
w_{p_{n-1,l}^{n,k}}(\rho_{n-1,l}-\rho_{n,k})} {\prod_{j,l=1\atop
(j\ne l)}^{n-2} w_{p_{n-1,j}^{n-1,l}}(\rho_{n-1,j}-\rho_{n-1,l})}
\prod_{l=1}^{n-2} \frac{w_{p_{n-1,l}}(-\rho_{n-1,l})}{w_{\tilde
p_{n-1,l}}(\rho_{n-1,l})}\eea
is eigenvector of $\;B_n(\la)$:\\ [-6mm]
\be\label{Blm}
B_n(\lm)\Psi_{\bdr_n} =\lm\, r_{n,0}\om^{-\rho_{n,0}}
\prod_{k=1}^{n-1}\lk\lm+r_{n,k}\om^{-\rho_{n,k}}\rk\Psi_{\bdr_n}.
\ee
The Fermat curve points $\;\tp_n,\:p_{n,l},\:\tp_{n,l},\:p_{n',l}^{n,k}$
and $\:r_{n,k},\:$ entering (\ref{QQQ})
are related to the parameters of the model $\:a_s,b_s,c_s,d_s,\varkappa_s\:$
by equations (\ref{xxx}), \r{rel_other}, \r{xynull}, \r{xyt}.

$A_n(\lm)$ acts on $\Psi_{\bdr_n}$ as follows:
\bea
\lefteqn{A_n(\lm)\Psi_{\bdr_n}\:=\:\prod_{s=1}^{n-1} \lk
1-\frac{\lm}{\lm_{n,s}}\rk\:\Psi_{\bdr_n} +\lm \kp_1\cdots\kp_n\:
\prod_{s=1}^{n-1} (\lm-\lm_{n,s})\,\Psi_{\bdr_n^{+0}}\;+}\ny\\ &&
\hs\hs\hs+\;\sum_{k=1}^{n-1}\; \lk\prod_{s\ne k}
\frac{\lm-\lm_{n,s}}{\lm_{n,k}-\lm_{n,s}}\!\rk\!
\frac{\lm}{\lm_{n,k}}\;\varphi_k(\bdr'_n)\;\Psi_{\bdr_n^{+k}},
\label{Alm}
\eea
where\\ [-11mm]
\be \label{phik} \varphi_k(\bdr_n')\;=\;-\frac{\tilde r_{n-1}}{r_n}\;\om^{-\tilde
\rho_{n}+\rho_{n,0}}\;F_n(\lm_{n,k}/\om)\;\prod_{s=1}^{n-2}y_{n-1,s}^{n,k}\,
 \ee
 with\\ [-10mm]
\be F_n(\lm)\:=\:\lk\, b_n\,+\om a_n \,\kp_n \lm\rk\,
\lk \,\la\, c_n\,+d_n/\kp_n\, \rk. \label{qdet}\ee
{\it\bf Corollary.}
In particular, at the $\:n-1\:$ zeros $\:\la_{n,k}\:$ of the
eigenvalue polynomial of $\;B_n(\la)\;$
\be \la_{n,k}\:=\:-r_{n,k}\om^{-\rho_{n,k}},\hs\hs k=1,\ldots,n-1\ee
the operator $\:A_n\:$ acts as shift operator for the $k$-th index of $\;\Psi_{\rho_n}$:
\be
 A_n\lk\lm_{n,k}\rk\Psi_{\bdr_n}\,=\, \varphi_k(\bdr_n')\;\Psi_{\bdr_n^{+k}}\,.\label{Almk}
 \ee
Further, the term in \r{Alm} of highest degree in $\lm$ gives:
$\bV_n=\bv_1\bv_2\ldots\bv_n\;$ is a shift operator for the zeroth index of $\Psi_{\bdr_n}$:
\be\label{vvvPsi} \bV_n\,\Psi_{\bdr_n}\,=\,\Psi_{\bdr_n^{+0}}\,. \ee
\end{theorem}
\noindent {\it\bf Proof.}
We shall prove the Theorem~\ref{th1}  by induction,
showing that if it is valid for $n-1$ site eigenvectors, then it
follows for $n$ site eigenvectors. Namely,
we assume the correctness of the following formulas
\bea\label{Blm1}
B_{n-1}(\lm)\Psi_{\bdr_{n-1}} &=& \lm\, r_{n-1,0}\om^{-\rho_{n-1,0}}
\prod_{l=1}^{n-2}\lk\lm-\lm_{n-1,l}\rk\Psi_{\bdr_{n-1}},\\
A_{n-1}(\lm)\Psi_{\bdr_{n-1}}&=&\sum_{l=1}^{n-2} \left(\prod_{s\ne l}
\frac{\lm-\lm_{n-1,s}}{\lm_{n-1,l}-\lm_{n-1,s}}\right)
\frac{\lm}{\lm_{n-1,l}}\,\varphi_l(\bdr'_{n-1}) \Psi_{\bdr_{n-1}^{+l}}+
     \ny\\  \label{Alm1}
     &&\hspace*{-27mm} +\;\prod_{s=1}^{n-2} \left(1-\frac{\lm}
     {\lm_{n-1,s}}\right)\Psi_{\bdr_{n-1}}
     +\lm \kp_1\cdots\kp_{n-1} \prod_{s=1}^{n-2}
     (\lm-\lm_{n-1,s})\cdot \Psi_{\bdr_{n-1}^{+0}},
\eea
where
$\lm_{n-1,l}\,=\,-r_{n-1,l}\om^{-\rho_{n-1,l}}$ and the formulas for
$\varphi_l(\bdr_{n-1}')$ given by (\ref{phik}) with $n$
replaced by $n-1$.

\vspace{2mm}

\noindent
{\sf Formula \r{Blm} for $\:B_n(\lm)\Psi_{\bdr_n}$}:\\ [2mm]
In order to prove the eigenvalue formula (\ref{Blm}) we use the following relation
\be\label{recB}
B_n(\lm)=A_{n-1}(\lm)\, \lm\,\bu_n^{-1} (a_n-b_n \bv_n)+
B_{n-1}(\lm) \left(\lm a_n c_n+\frac{b_n d_n}{\kp_n} \bv_n\right)
\ee
which follows directly from \r{bazh_strog} and \r{mm}. We apply its left-hand
side to the left-hand side of \r{PSI} and its right-hand side to the
right-hand side of \r{PSI}. On the right, we use \r{Blm1}, \r{Alm1}, \r{efb}, \r{vpsi}.
According to \r{Alm1}, $A_{n-1}$
introduces shifts in the indices $\bdr_{n-1}$ of $\:\Psi_{\bdr_{n-1}}\,$, while the
second term involving $\bv_n$ shift the index of $\:\psi^{(n)}_{\rho_n}$. Since we are
looking for an eigenstate, by shifting the summation indices we restore the original
indices. However, this leaves a change in the matrix $\:Q(\bdr_{n-1},\rho_n|\bdr_n)\,$.
Now the difference equation \r{Fermat} for the $w_p(\g)$ functions appearing in
$\:Q(\bdr_{n-1},\rho_n|\bdr_n)\,$ is used, producing several factors under the summation
which together we call $\;R\,$:
\[
B_n(\lm)\Psi_{\bdr_n}=\sum_{\bdr_{n-1}\in (\ZN)^{n-1}\atop \rho_n \in \ZN}
 Q(\bdr_{n-1},\rho_n|\bdr_n)\, R\, \Psi_{\bdr_{n-1}}\otimes \psi^{(n)}_{\rho_n},
\]
After some calculation we obtain
\[\fl
R\;=\;\left\{\sum_{l=1}^{n-2} \left(\prod_{s\ne l}
\frac{\lm-\lm_{n-1,s}}{\om\lm_{n-1,l}-\lm_{n-1,s}}\right)
\frac{\lm}{\om\lm_{n-1,l}}\;\varphi_l({\bdr'}^{-l}_{n-1})\,
\frac{Q(\bdr^{-l}_{n-1},\rho_n|\bdr_n)}{Q(\bdr_{n-1},\rho_n|\bdr_n)}\;\:+\:\right.
\]
\[\fl\qquad\left.
+\prod_{s=1}^{n-2} \left(1-\frac{\lm}{\lm_{n-1,s}}\right)
+\lm \kp_1\cdots\kp_{n-1} \prod_{s=1}^{n-2} (\lm-\lm_{n-1,s})
\frac{Q(\bdr^{-0}_{n-1},\rho_n|\bdr_n)}{Q(\bdr_{n-1},\rho_n|\bdr_n)}\right\}
\lm r_n\om^{-\rho_n}\;+
\]
\[
+ \lm r_{n-1,0}\om^{-\rho_{n-1,0}}
\prod_{l=1}^{n-2}\lk\lm-\lm_{n-1,l}\rk
\left(\lm a_n c_n +\frac{b_n d_n}{\kp_n}\cdot
\frac{Q(\bdr_{n-1},\rho_n-1|\bdr_n)}{Q(\bdr_{n-1},\rho_n|\bdr_n)}\right).
\]
We have to show that\\ [-8mm]
\be\label{R}
R=\lm\, r_{n,0}\om^{-\rho_{n,0}}
\prod_{k=1}^{n-1}(\lm-\lm_{n,k});\qquad \lm_{n,k}=-r_{n,k}\om^{-\rho_{n,k}}.
\ee
This  will prove (\ref{Blm}).
Using the definitions (\ref{xxx}) of $\:x_{m,s}\:$ and $\:\tilde x_{m,s}\:$,
we can rewrite \r{qdet} for the argument $\lm=\lm_{n,k}/\om$ as follows:
\be F_n(\lm_{n,k}/\om)\:=\:\lm_{n,k}b_n c_n\om^{-1}\lk 1-x_{n,k}\om^{-\rho_{n,k}}\rk\:
\lk 1-\tilde x_{n,k}\om^{\rho_{n,k}+1}\rk.\label{qdetmk}\ee
Taking into account the expression (\ref{QQQ}) for $Q(\bdr_{n-1},\rho_n|\bdr_n)$,
the definition for $w_p(\g)$ and the relations
\r{xxx},\r{rel_other},\r{rtilde},\r{xynull},\r{qdetmk} we obtain:
\[\fl
\frac{Q(\bdr^{-l}_{n-1},\rho_n|\bdr_n)}{Q(\bdr_{n-1},\rho_n|\bdr_n)}=
\om^{\rho_n-\rho_{n,0}}
\prod_{k=1}^{n-1} \frac{w_{p^{n,k}_{n-1,l}}(\rho_{n-1,l}-\rho_{n,k}-1)}
{w_{p^{n,k}_{n-1,l}}(\rho_{n-1,l}-\rho_{n,k})}
\cdot
\frac{w_{p_{n-1,l}}(-\rho_{n-1,l}+1)} {w_{p_{n-1,l}}(-\rho_{n-1,l})}
\times\]
\[\fl\;\times
\frac{w_{\tp_{n-1,l}}(\rho_{n-1,l})}{w_{\tp_{n-1,l}}(\rho_{n-1,l}-1)}\,
\prod_{s\ne l}\left(
\frac{w_{p^{n-1,l}_{n-1,s}}(\rho_{n-1,s}-\rho_{n-1,l})}
{w_{p^{n-1,l}_{n-1,s}}(\rho_{n-1,s}-\rho_{n-1,l}+1)}
\cdot
\frac{w_{p^{n-1,s}_{n-1,l}}(\rho_{n-1,l}-\rho_{n-1,s})}
{w_{p^{n-1,s}_{n-1,l}}(\rho_{n-1,l}-\rho_{n-1,s}-1)}
\right)=
\]\[
=\frac{\om}
{\varphi_l({\bdr'}^{-l}_{n-1})}
\frac{r_{n,0}\,\om^{-\rho_{n,0}}} {r_n\om^{-\rho_n}}
\prod_{k=1}^{n-1} (\lm_{n-1,l}-\lm_{n,k})
\prod_{s\ne l}\frac{\om\lm_{n-1,l}-\lm_{n-1,s}}{\lm_{n-1,l}-\lm_{n-1,s}},
\]
\[\fl
\frac{Q(\bdr^{-0}_{n-1},\rho_n|\bdr_n)}{Q(\bdr_{n-1},\rho_n|\bdr_n)}=
\om^{\rho_n-\rho_{n,0}}
\frac{w_{p_{n0}}(\rho_{n,0}-\rho_{n-1,0}-1)}
{w_{p_{n0}}(\rho_{n,0}-\rho_{n-1,0})}=
\frac{r_{n,0}\om^{-\rho_{n,0}}-r_{n-1,0}\om^{-\rho_{n-1,0}}a_n c_n}
{\varkappa_1 \varkappa_2\cdots \varkappa_{n-1} r_n\om^{-\rho_n}},
\]
\[\fl
\frac{Q(\bdr_{n-1},\rho_n-1|\bdr_n)}{Q(\bdr_{n-1},\rho_n|\bdr_n)}=
\om^{\trh_{n-1}-\trh_n} \frac{w_{\tp_n}(\trh_n-\rho_n-1)}{w_{\tp_n}(\trh_n-\rho_n)}=
\frac{\varkappa_n}{b_n d_n}\cdot
\frac{\rt_n\om^{-\trh_n}-r_n \om^{-\rho_n}}
{\rt_{n-1}\om^{-\trh_{n-1}}}.
\]
Substituting these expressions into $R$ gives
\[\fl
R\;=\;\left\{\sum_{l=1}^{n-2} \left(\prod_{s\ne l}
\frac{\lm-\lm_{n-1,s}}{\lm_{n-1,l}-\lm_{n-1,s}}\right)
\frac{\lm}{\lm_{n-1,l}}
\frac{r_{n,0}\,\om^{-\rho_{n,0}}} {r_n\om^{-\rho_n}}
\prod_{k=1}^{n-1} (\lm_{n-1,l}-\lm_{n,k})+\right.
\]
\[\fl\quad\left.
+\prod_{s=1}^{n-2} \left(1-\frac{\lm}{\lm_{n-1,s}}\right)
+\lm\prod_{s=1}^{n-2} (\lm-\lm_{n-1,s})\cdot
\frac{r_{n,0}\om^{-\rho_{n,0}}-r_{n-1,0}\om^{-\rho_{n-1,0}}a_n c_n}
{ r_n\om^{-\rho_n}}\right\} \lm r_n\om^{-\rho_n}+
\]
\[
+ \lm r_{n-1,0}\om^{-\rho_{n-1,0}}
\prod_{l=1}^{n-2}\lk\lm-\lm_{n-1,l}\rk\cdot
\left(\lm a_n c_n +\frac{\rt_n\om^{-\trh_n}-r_n \om^{-\rho_n}}
{\rt_{n-1}\om^{-\trh_{n-1}}}\right).
\]
After appropriate cancellations this becomes
\[\fl
R\;=\;\sum_{l=1}^{n-2} \left(\prod_{s\ne l}
\frac{\lm-\lm_{n-1,s}}{\lm_{n-1,l}-\lm_{n-1,s}}\right)
\frac{\lm^2}{\lm_{n-1,l}}
r_{n,0}\,\om^{-\rho_{n,0}}
\prod_{k=1}^{n-1} (\lm_{n-1,l}-\lm_{n,k})\:+
\]
\be\label{RR}
\fl\qquad\qquad\quad +\:\lm^2\prod_{s=1}^{n-2} (\lm-\lm_{n-1,s})\cdot
r_{n,0}\om^{-\rho_{n,0}}\,+ \lm\, \rt_n\om^{-\trh_n}
 \prod_{l=1}^{n-2}\left(1-\frac{\lm}{\lm_{n-1,l}}\right)\,.
\ee
To prove (\ref{R}) we note that the coefficients at $\lm^n$ in both
expressions (\ref{R}) and (\ref{RR}) are $r_{n,0}\om^{-\rho_{n,0}}$
and coefficients at $\lm$ also coincide being
$\rt_n\om^{-\trh_n}$.
Therefore the difference of these two expressions
has the form $\lm^2 P(\lm)$ where $P(\lm)$ is a polynomial of degree $n-3$.
Using the explicit expressions (\ref{R}) and (\ref{RR}) we convince
ourselves that $P(\lm_{n-1,j})=0$ and therefore $P(\lm)\equiv 0$.
This completes the proof that $\Psi_{\bdr_n}$ defined by \r{PSI}, (\ref{QQQ})
is eigenvector of $B_n(\lm)$ with eigenvalue (\ref{R}).

\vspace{2mm}

\noindent
{\sf Formula \r{Almk} for $\:A_n(\lm_{n,k})\Psi_{\bdr_n}$}: \\ [2mm]
Next we show the validity of (\ref{Almk}), (\ref{phik}).
We will need the relation
\be \fl \,\bu_n^{-1} (a_n-b_n \bv_n)
A_n(\lm)\;=\;(1+\lm\kp_n\om^{-1}\bv_n) B_n(\lm)/\lm
         \,-\,\bv_n\,F_n(\lm/\om)\:B_{n-1}(\lm)/\lm \label{recAB} \ee
which can be obtained by eliminating
$\:A_{n-1}\:$ between \r{recB} and
\be   A_n(\lm)=(1+\lm\kp_n\bv_n) A_{n-1}(\lm)+\bu_n(c_n-d_n \bv_n)
                             B_{n-1}(\lm).\label{recA}\ee
Let us apply (\ref{recAB}) to
$\:\Psi_{\bdr_n}\:$ for $\la=-r_{n,k}\om^{-\rho_{n,k}}$,
i.e. at the zeros of $\:B_n(\la)$.
This gives
\bea \fl \bu_n^{-1} (a_n-b_n \bv_n) A_n(\lm_{n,k}) \Psi_{\bdr_n}&=&
-F_n(\lm_{n,k}/\om)/\lm_{n,k}\;\times\ny\\ [1mm]
\fl &\times&\!\!\! \sum_{\bdr_{n-1}\in (\ZN)^{n-1}\atop \rho_n \in
\ZN}\!\!
 Q(\bdr_{n-1},\rho_n|\bdr_n)\; B_{n-1}(\lm_{n,k})\Psi_{\bdr_{n-1}}\otimes
\psi^{(n)}_{\rho_n+1}.\label{ABm} \eea
From (\ref{Blm1}) we know how to apply
$B_{n-1}$ to $\Psi_{\bdr_{n-1}}$:
\bea\label{Blmnk}
\fl \lefteqn{B_{n-1}(\lnk)\Psi_{\bdr_{n-1}} =\la_{n,k}\,
r_{n-1,0}\om^{-\rho_{n-1,0}} \prod_{s=1}^{n-2}\lk
-\rnk\om^{-\rho_{n,k}}+r_{n-1,s}\om^{-\rho_{n-1,s}}\rk\Psi_{\bdr_{n-1}}}\ny\\ &=&
\la_{n,k}\,
\rt_{n-1}\om^{-\trh_{n-1}} \lk\prod_{s=1}^{n-2}y_{n-1,s}^{n,k}\:
   \frac{w_{p_{n-1,s}^{n,k}}\lk\rho_{n-1,s}-\rho_{n,k}-1\rk}
{w_{p_{n-1,s}^{n,k}}\lk\rho_{n-1,s}-\rho_{n,k}\rk}\rk\: \Psi_{\bdr_{n-1}}\,.
\eea
Using (\ref{efb}) we find the action of the inverse of the operator
$\bu_n^{-1} (a_n-b_n \bv_n)$ on $\psi^{(n)}_{\rho_n}$\,:
\be\label{invpsi}
(\bu_n^{-1} (a_n-b_n \bv_n))^{-1} \psi^{(n)}_{\rho_n}=
r_n^{-1} \om^{\rho_n} \psi^{(n)}_{\rho_n}.
\ee
Shifting the summation over $\rho_n$ in \r{ABm} and then applying
(\ref{invpsi}) we obtain
\bea  \fl A_n(\lm_{n,k}) \Psi_{\bdr_n}&=&
-r_n^{-1} F_n(\lm_{n,k}/\om)/\lm_{n,k}  \;\times\ny\\ [1mm]
&&\hspace*{-15mm}\times\!\!\!\sum_{\bdr_{n-1}\in (\ZN)^{n-1}\atop \rho_n \in
\ZN}
 Q(\bdr_{n-1},\rho_n-1|\bdr_n)\,\om^{\rho_n}
 \; B_{n-1}(\lm_{n,k})\:\Psi_{\bdr_{n-1}}\otimes \psi^{(n)}_{\rho_n}. \eea
Finally, using (\ref{Blmnk}) and observing that
\[\fl
Q(\bdr_{n-1},\rho_n\!-1|\bdr_n)\:\om^{\rho_n\!-\trh_{n-1}}\,
\prod_{s=1}^{n-2}\frac{w_{p_{n-1,s}^{n,k}}
\lk\rho_{n-1,s}\!-\rho_{n,k}-1\rk}
{w_{p_{n-1,s}^{n,k}}\lk\rho_{n-1,s}\!-\rho_{n,k}\rk}\,=\,\om^{-\trh_{n}+\rho_{n,0}}
\;Q(\bdr_{n-1},\rho_n|\bdr_n^{+k}) \]
we come to (\ref{Almk}).

\vspace{2mm}

\noindent
{\sf Formula \r{vvvPsi} for $\;\bV_n\Psi_{\bdr_n}$}:\\ [2mm]
Using $\:\bV_{n-1} \Psi_{\bdr_{n-1}}\,=\,\Psi_{\bdr_{n-1}^{+0}}$
and $\:\bv_{n} \psi^{(n)}_{\rho_n}\,=\,\psi^{(n)}_{\rho_n+1}\:$  we have
\bea\fl\lefteqn{\bV_{n-1}\,\bv_{n}\Psi_{\bdr_{n}}=
\!\!\!\!\sum_{\bdr_{n-1}\in(\ZN)^{n-1}\atop \rho_{n} \in \ZN}
 \!\!\!Q(\bdr_{n-1},\rho_{n}|\bdr_{n})
 \Psi_{\bdr_{n-1}^{+0}}\otimes\,\psi^{(n)}_{\rho_n+1}}\ny\\ [-6.5mm]
&& \hspace*{3cm}=\!\!\!\!\sum_{\bdr_{n-1}\in(\ZN)^{n-1}\atop \rho_n \in \ZN}
 \!\!Q(\bdr_{n-1}^{-0},\rho_n-1|\bdr_n) \Psi_{\bdr_{n-1}}
 \otimes\,\psi^{(n)}_{\rho_n}, \eea
where in the second line we have shifted the summation variables
$\rho_{n-1,0}$ and $\rho_n$. Now considering the explicit form
(\ref{QQQ}) for $\:Q(\bdr_{n-1},\rho_n|\bdr_n)\:$ we read off that
\be
Q(\bdr_{n-1}^{-0},\rho_n-1|\bdr_n)\:=\:Q(\bdr_{n-1},\rho_n|\bdr_n^{+0})\hs
\ee
which gives (\ref{vvvPsi}).

\vspace{2mm}

\noindent
{\sf Formula \r{Alm} for $A_n(\lm)\Psi_{\bdr_n}$}:\\ [2mm]
The operator $A_n(\la)$ is a polynomial in $\la$ of $n$th order.
From \r{bazh_strog} and \r{mm} we immediately read off its
the highest and lowest coefficients:   \be
A_n(\la)\:=\:1\:+\:\ldots\:+\:\la^n\kp_1\kp_2\ldots\kp_n\,\bV_n\,.
\label{An}  \ee
Using \r{vvvPsi} we know how these terms act on $\Psi_{\bdr_n}$
and if in addition we use the action of $\:A_n\,$ at the $n-1$ particular
values given in \r{Almk}, we can
reconstruct the action of the whole
polynomial $A_n(\la)$ on $\:\Psi_{\bdr_n}\:$ uniquely.
Comparing this to \r{Alm} we see that formula \r{Alm} is the one which
satisfies all these data.
Therefore by uniqueness formula (\ref{Alm}) is one which we are looking for.

This completes the proof of Theorem~\ref{th1}.

\section{Action of $D_{n}$ on the eigenstates of $B_{n}$.}

In order to obtain the action of $\,D_n(\lm)\,$ on $\,\Psi_{\bdr_n}$ we use the notion of
the quantum determinant ${\det}_q\, T_n(\lm)$ of the monodromy matrix.
Since the rank of the matrix $R(\om\lm,\lm)$ is $1$, the definition of the quantum determinant
is given by
\be\label{qdett}\fl
R(\om\lm,\lm)\; T_n^{(1)}(\om\lm)\,T_n^{(2)}(\lm)\,=
T_n^{(2)}(\lm)\, T_n^{(1)}(\om\lm)\, R(\om\lm,\lm)\,=:
{\det}_q\, T_n(\lm) \cdot\, R(\om\lm,\lm).
\ee
Explicitly we have
\be\label{qdetexpl}
{\det}_q\, T_n(\lm)=A_n(\om \lm)D_n(\lm)-C_n(\om \lm) B_n(\lm).
\ee
Using (\ref{mm}) and (\ref{qdett}) we obtain
the factorization property of the quantum determinant
\[ {\det}_q\, T_n(\lm)={\det}_q\, L_1(\lm)\cdot
{\det}_q\, L_2(\lm)\cdots\:{\det}_q\,L_n(\lm), \]
For a single $L$-operator, using \r{qdet}, \r{qdetexpl} gives
$\;{\det}_q\,L_m(\lm)\:=\:\bv_m\,F_m(\lm)\,$. So
\be  A_n(\om \lm)D_n(\lm)-C_n(\om \lm) B_n(\lm)
       \:=\:\bV_n\cdot\prod_{m=1}^n\;F_m(\lm).\label{qdetall}
       \ee
Acting by both sides of this identity on $\Psi_{\bdr_n},$
fixing $\lm=\lm_{n,k}$ (i.e. at the zeroes of the eigenvalue polynomial of $B_n(\lm)$)
and using the inverse of
(\ref{Almk}) with (\ref{phik}), we
see that, very similar to $\:A_n(\la_{n,k}),$ also $\:D_n(\la_{n,k})\:$
acts as a shift operator on $\Psi_{\bdr_n}$, compare \r{dlm}:
\be\fl
D_n(\lm_{n,k})\Psi_{\bdr_n}=\tilde\varphi_k(\bdr_n')\,\Psi_{\bdr_n^{+0,-k}};\qquad
 \tilde\varphi_k(\bdr'_n)=
-\frac{r_n}{\tilde r_{n-1}}\frac{\om^{\trh_n-\rho_{n,0}-1}}
{\prod_{s=1}^{n-2}y_{n-1,s}^{n,k}}\;\prod_{m=1}^{n-1}\;F_m(\lm_{n,k}). \label{Dlmk}
\ee
Note that $D_n(\lm_{n,k})$ shifts $\rho_{n,k}$ in the opposite
direction as $A_n(\lm_{n,k})$ (see \r{phik} and \r{Almk})
and $D_n(\lm_{n,k})$ also
shifts $\rho_{n,0}$. The shift in $\rho_{n,0}$
is due to the operator $\bV_n$ at the right-hand side of (\ref{qdetall}).
Apart from the shifts just mentioned, applying the inverse of $A_n(\om \la)$ has cancelled
in \r{Dlmk} the last factor $m=n$ of the quantum determinant (\ref{qdetall}).
Analogously to (\ref{Alm}), using the particular values (\ref{Dlmk}) and reading off
the coefficients of $\la^0$ and $\la^n$ directly from
(\ref{mm}), we obtain the following interpolation formula
for $D_n(\lm)\Psi_{\bdr_{n}}$:
\bea
\fl D_n(\lm)\Psi_{\bdr_{n}}&=&\sum_{k=1}^{n-1} \lk\prod_{s\ne k}
\frac{\lm-\lm_{n,s}}{\lm_{n,k}-\lm_{n,s}}\rk
\frac{\lm}{\lm_{n,k}}\,\tilde\varphi_k(\bdr'_{n})\Psi_{\bdr_{n}^{+0,-k}}+\ny\\
\fl &+&\prod_{s=1}^{n-1} \lk 1-\frac{\lm}{\lm_{n,s}}\rk\cdot
\prod_{m=1}^n\frac{b_m d_m}{\kp_m} \cdot
\Psi_{\bdr_{n}^{+0}} +\lm \prod_{m=1}^n a_m
c_m \cdot \prod_{s=1}^{n-1} (\lm-\lm_{n,s})\cdot \Psi_{\bdr_{n}}.
\label{Dlm}\eea

\section{Periodic model. Baxter equation and functional relations}
\label{periodic}
\subsection{The Baxter equations}

After having determined the eigenvalues and eigenvectors of the auxiliary system, we
now perform the first step of the program exposed in Subsection \ref{SoV}, i.e.
the calculation of the eigenvalues and eigenvectors of the inhomogenous
$n$-site {\it periodic} \BBS chain model with the transfer matrix \r{tr_matl},
\r{tr_mat_bs}. Following the ideas  of the papers
\cite{Gu81,Skly1,Sk85,Skly2,KarLeb1} we are looking for eigenvectors of ${\bf t}_n(\lm)$ as
linear combinations of the eigenvectors $\Psi_{{\bdr}_{n}}$ of the auxiliary system.

It is convenient to go by Fourier transform in $\rho_{n,0}$
from $\Psi_{\bdr_n}$ to a basis of eigenvectors of $\bV_n$
(and therefore of the Hamiltonians $\:{\bf H}_0$ and $\:{\bf H}_n$, see \r{tr_mat_bs}, \r{Znc})
\be \tilde\Psi_{\rho,\bdr'_n}\:=\:
{\textstyle \sum_{\rho_{n,0}\in\ZN}}\;\om^{-\rho\cdot\rho_{n,0}}\:
     \Psi_{\bdr_n};\qquad\bV_n \:\tilde \Psi_{\rho,\bdr'_n}\;=\;\om^\rho\: \tilde
     \Psi_{\rho,\bdr'_n}.\ee
A shift of $\,\rho_{n,0}\,$ in $\Psi_{\bdr_n}$ is replaced by a multiplication of $\tilde
\Psi_{\rho,\bdr'_n}$ by powers of $\om$.
So from (\ref{Alm}) and (\ref{Dlm}), the action of $\:{\bf t}_n(\la)$ on
$\tilde\Psi_{\rho,\bdr'_{n}}$ becomes
\bea \fl \lefteqn{ {\bf t}_n(\la)\tilde\Psi_{\rho,{\bdr'}_{n}}
=\,\sum_{k=1}^{n-1}\lk\prod_{s\ne k} \frac{\lm\!-\lm_{n,s}}{\lm_{n,k}\!-\lm_{n,s}}\rk
\!\frac{\lm}{\lm_{n,k}}\lk\varphi_k(\bdr'_{n})\tilde\Psi_{\rho,{\bdr'}_{n}^{+k}}\!
           +\om^\rho \tilde\varphi_k(\bdr'_{n})\tilde\Psi_{\rho,{\bdr'}_{n}^{-k}}\rk} \ny\\
\fl &&\hspace*{30mm}+\:
\lb\lk 1+\om^\rho\prod_{m=1}^n\frac{b_m d_m}{\kp_m}\rk\:
   \prod_{s=1}^{n-1} \lk 1-\frac{\la}{\la_{n,s}}\rk\,+\right.\ny\\
\fl && \hspace*{50mm}\left.+ \:\la \lk \om^\rho\,\prod_{m=1}^n\:\kp_m
+\prod_{m=1}^n\: a_mc_m \rk
\prod_{s=1}^{n-1}(\lm-\lm_{n,s}) \rb
\tilde\Psi_{\rho,\bdr'_{n}},
\label{ApD}\eea
where we have taken into account that
$\varphi_k(\bdr'_{n})$ and $\tilde\varphi_k(\bdr'_{n})$ are
independent of $\rho_{n,0}$.
Of course, since $\:{\bf t}_n(\la)\,$ commutes with $\bV_n$, in (\ref{ApD}) there is
no coupling between sectors of different $\rho$ and we
get separate equations for the different ``charge'' quantum numbers $\rho$ which
often will not be indicated explicitly.

Let $\Phi_{\rho,{\bf E}}$ be eigenvector of ${\bf t}_n(\lm)$ with eigenvalue
\be  t_n(\lm|\,\rho,\,{\bf E})\;=\;
E_0+E_1\lm+\cdots+E_{n-1}\lm^{n-1}+E_{n}\lm^{n}\,, \label{tnh}\ee
where $\;\:{\bf E}\;=\;\{E_1,\ldots,E_{n-1}\}\;$ and from \r{Znc}
the values of $E_0$ and $E_n$
are
\be\label{iE0En}
 E_0\,=1+\om^\rho \prod_{m=1}^n\frac{b_m d_m}{\kp_m},\qquad
E_n\,=\prod_{m=1}^n a_m c_m\,+\om^{\rho}\,\prod_{m=1}^n\kp_m\,.\ee
We are looking for $\:\Phi_{\rho,{\bf E}}\:$ to be of the form
\be \Phi_{\rho,{\bf E}}\:=\:
\sum_{\bdr'_n} \; Q(\bdr'_{n}|\,\rho,{\bf E})\;\:\tilde\Psi_{\rho,{\bdr'}_{n}}.
\label{EV}\ee
From (\ref{ApD}) we get a difference equation for $\:Q (\bdr'_{n}|\,\rho,{\bf E})\:$
with respect to variables
$\:\bdr'_{n}\:$ which depends on $\;\lm$:
\bea \fl
t_n(\lm|\rho,{\bf E})\:Q (\bdr'_{n}|\,\rho,{\bf E})\:&=&\:\sum_{k=1}^{n-1}
\frac{\lm}{\om\lm_{n,k}}\,\varphi_k({\bdr'}^{-k}_{n})\;
Q({\bdr'}^{-k}_{n}|\,\rho,{\bf E})
        \prod_{s\ne k}\frac{\lm-\lm_{n,s}}{\om\lm_{n,k}-\lm_{n,s}}\:\: +
\ny\\  &&\hspace*{-10mm}+\:
\sum_{k=1}^{n-1}\frac{\om\lm}{\la_{n,k}}\,\om^\rho
\tilde\varphi_k({\bdr'}^{+k}_{n})\; Q({\bdr'}^{+k}_{n}|\,\rho,{\bf E})
\prod_{s\ne k}\frac{\lm-\lm_{n,s}}{\om^{-1}\la_{n,k}-\la_{n,s}}\,+\ny\\
 &&\hspace*{-10mm}+\:
\lb\lk 1+\om^\rho\prod_{m=1}^n\frac{b_m d_m}{\kp_m}\rk
\prod_{s=1}^{n-1} \lk 1-\frac{\la}{\la_{n,s}}\rk\:+\right.\ny\\
 &&\hspace*{-3mm}  \left.+\,\la
 \lk \om^\rho\prod_{m=1}^n \kp_m+\prod_{m=1}^n a_mc_m \rk
\prod_{s=1}^{n-1}(\la-\la_{n,s}) \rb Q(\bdr'_n|\,\rho,{\bf E}).
\eea

Substituting sequentially $\lm=\lm_{n,k}$, $k=1,2,\ldots,n-1$, we
obtain a system of difference equations with respect to the variables
$\bdr'_{n}$:
\bea  \fl \lefteqn{ t_n(\lm_{n,k}|\rho,\,{\bf E})\:\:
Q (\bdr'_{n}|\,\rho,{\bf E})=
\lk\prod_{s\ne k} \frac{\lm_{n,k}-\lm_{n,s}}{\om\lm_{n,k}-\lm_{n,s}}\rk
\om^{-1}\,\varphi_k({\bdr'}^{-k}_{n})\: Q ({\bdr'}^{-k}_{n}|\,\rho,{\bf E})\;+}\ny\\
\fl &&\hspace*{10mm}+\:\lk\prod_{s\ne k}
\frac{\lm_{n,k}-\lm_{n,s}}{\om^{-1}\lm_{n,k}-\lm_{n,s}}\rk
\om^{\rho+1} \tilde\varphi_k({\bdr'}^{+k}_{n})\: Q({\bdr'}^{+k}_{n}|\,\rho,{\bf E}),
\hq k=1,\ldots,n-1.
\label{pdiff}\eea
In analogy to \cite{Gu81,Sk85,KarLeb1} we can decouple these difference equations
using a Sklyanin's measure, namely, by
 introducing  $\;\tQ (\bdr'_{n}|\,\rho,{\bf E})\;$ defined as
\be Q (\bdr'_{n}|\,\rho,{\bf E})=\frac{\tQ (\bdr'_{n}|\,\rho,{\bf E})}
{\prod_{s,s'=1\atop (s\ne s')}^{n-1}
w_{p_{n,s}^{n,s'}}(\rho_{n,s}-\rho_{n,s'})}.\ee
Rewriting \r{pdiff} in terms of $\tQ$ produces factors $R_\pm$
in both terms of the right hand side:
\bea  \fl \lefteqn{ t_n(\lm_{n,k}|\rho,\,{\bf E})\:\:
\tQ (\bdr'_{n}|\,\rho,{\bf E})=
\lk\prod_{s\ne k} \frac{\lm_{n,k}-\lm_{n,s}}{\om\lm_{n,k}-\lm_{n,s}}\rk
\om^{-1}\,\varphi_k({\bdr'}^{-k}_{n})\:R_-\:\tQ({\bdr'}^{-k}_{n}|\,\rho,{\bf E})\;+}\ny\\
\fl &&\hspace*{10mm}+\:\lk\prod_{s\ne k}
\frac{\lm_{n,k}-\lm_{n,s}}{\om^{-1}\lm_{n,k}-\lm_{n,s}}\rk
\om^{\rho+1} \tilde\varphi_k({\bdr'}^{+k}_{n})\:R_+\:\tQ({\bdr'}^{+k}_{n}|\,\rho,{\bf E}),
\hq k=1,\ldots,n-1.
\label{pdiffR}\eea
where \bea \fl R_+\;&=&\prod_{s=1\atop{s\neq k}}^{n-1}\;
  \frac{w_{p_{n,s}^{n,k}}(\rho_{n,s}-\rho_{n,k})}
  {w_{p_{n,s}^{n,k}}(\rho_{n,s}-\rho_{n,k}\,- 1)}\:\cdot\:
  \frac{w_{p_{n,k}^{n,s}}(\rho_{n,k}-\rho_{n,s})}
     {w_{p_{n,k}^{n,s}}(\rho_{n,k}-\rho_{n,s}+ 1)}\ny\\
  \fl \;&=&\prod_{s=1\atop{s\neq k}}^{n-1}\;\frac{y_{n,s}^{n,k}}{1\,-x_{n,s}^{n,k}\:
  \om^{\rho_{n,s}-\rho_{n,k}}}\;
   \:\frac{1\,-x_{n,k}^{n,s}\:\om^{\rho_{n,k}-\rho_{n,s}+1}}{y_{n,k}^{n,s}}\;=\;
   \prod_{s=1\atop{s\neq k}}^{n-1}\;\:\frac{y_{n,s}^{n,k}}{y_{n,k}^{n,s}}\;
   \frac{\lm_{n,s}}{\lm_{n,k}}\;\:\frac{\om\,\lm_{n,s}-\lm_{n,k}}{\lm_{n,k}-\lm_{n,s}}
   \,, \ny\eea
and analogously $\:R_-$. We see that passing from $\:Q\:$ to $\:\tQ\:$ the brackets
containing differences of terms $\:\lm_{n,l}\:$ in \r{pdiffR}
are cancelled and so the equations decouple.
This means that in terms of $\;\tQ\,,\;$ the difference equations (\ref{pdiff}) admit the
{\it separation of variables}:
\be\tQ (\bdr'_{n}|\,{\bf E})\;=\;\prod_{k=1}^{n-1}\: \tilde
q_k(\rho_{n,k}).
\ee
Inserting the explicit expressions for $\varphi_k({\bdr'}^{-k}_{n})$
and $\tilde\varphi_k({\bdr'}^{+k}_{n})$
we obtain Baxter type difference equations for the functions $\qt_k(\rho_{n,k})$:
\be\fl
t_n(\lm_{n,k}|\rho,{\bf E})\;\qt_k(\rho_{n,k})\;=\;
\Delta_+(\lm_{n,k})\;\qt_k(\rho_{n,k}+1)\;+\;\Delta_-(\om\lm_{n,k})\;
\qt_k(\rho_{n,k}-1)\label{BAX}\ee
with\\ [-9mm]
\be \fl\Delta_+(\lm)\:=\:(\om^\rho/\chi_k)\,(\lm/\om)^{1-n}\:
         \prod_{m=1}^{n-1}\,F_m(\lm/\om);\qquad
\Delta_-(\lm)\:=\:\chi_k\:(\lm/\om)^{n-1}\:F_n(\lm/\om); \label{Dpm}\ee
\be
\chi_k\;=\;\frac{r_{n,0}\:\rt_{n-1}}{r_n\:\rt_n}\;\:(\prod_{s=1\atop s\ne
k}^{n-1}\; y_{n,k}^{n,s}/y^{n,k}_{n,s})\:\prod_{s=1}^{n-2}\:y^{n,k}_{n-1,s}\,.
\label{chi} \ee
In what follows we will mainly use $t(\lm)$ instead $t_n(\lm|\rho,{\bf E})$.
In fact the system of linear homogeneous equations (\ref{BAX}) with respect
to $\qt_k(\rho_{n,k})$, $\rho_{n,k}\in\ZN$, is not completely defined.
Since $E_1,$ $E_2,\ldots,$ $E_{n-1}$ are unknown, the coefficients $t(\lm_{n,k})$
are also unknown.
The requirement on the system of homogeneous equations (\ref{BAX}) for some
fixed $k$, $k=1,$ $2,\ldots,$ $n-1$,
to have a nontrivial solution leads to the requirement that the matrix of coefficients
must be degenerate.
The latter gives a relation for the values $E_0,$ $E_1,\ldots,$
$E_n$ entering $\:t(\lm)$.
Taking all such relations corresponding to all $k=1,$ $2,\ldots,$ $n-1$,
and using the values of $E_0$ and $E_n$ given in (\ref{iE0En}),
at least in principle we can find the
possible values of $E_1,\ldots,$ $E_{n-1}$. This fixes $\,t(\lm)$.
Then for every $k$, $k=1,$ $2,\ldots,$ $n-1$, we solve (\ref{BAX})
to find $\qt_k(\rho_{n,k})$ for $\rho_{n,k}\in\ZN\,$
(These difference equations have three terms and
cannot be solved in terms of the functions $w_p$).
This gives us finally  $Q(\bdr'_{n}|\,\rho,{\bf E})$ and therefore the
eigenvectors of the periodic \BBS model:
\[
\Phi_{\rho,{\bf E}}=
\sum_{\bdr_n=(\rho_{n,0},\ \bdr'_n)}
\om^{-\rho\cdot\rho_{n,0}}\, Q (\bdr'_{n}|\,\rho,{\bf E})\, \Psi_{\bdr_n}.
\]

\subsection{Role of the functional relations}
Now we will show that mentioned requirement on the systems of
homogeneous equations (\ref{BAX}) for all $k$ to have a nontrivial
solution is equivalent to functional relations \cite{BS,BBP,B_tau}
of the $\tau^{(2)}$-model. We define $\tau^{(0)}(\lm)=0$,
$\tau^{(1)}(\lm)=1$, $\tau^{(2)}(\lm)=t(\lm)$ (see \r{tnh}, \r{iE0En}) and
\be\fl \tau^{(j+1)}(\lm)=\tau^{(2)}(\om^{j-1}\lm)\,\tau^{(j)}(\lm)-
\om^{\rho}\,z(\om^{j-1}\lm)\, \tau^{(j-1)}(\lm),\qquad j=2,3,\ldots,
N, \label{rectau} \ee
where \\ [-11mm]
\be\label{zDelta}
 z(\lm)=\om^{-\rho} \Delta_+(\lm)\Delta_-(\lm)\:=\:\prod_{m=1}^{n} F_m(\lm/\om).
\ee
The appearance of the monodromy determinant (\ref{qdet}) in the fusion relation
is  a direct consequence of the fusion procedure (see \cite{KS82,KR87}).

The fusion hierarchy can be used to find eigenvalues of the transfer matrices
in lattice integrable models. A key tools here is, in addition to
(\ref{rectau}) to demand a ``truncation'' identity which allows to express $\tau^{(j)}(\lm)$
for some value $j$ through $\tau^{(i)}(\lm)$ with $i<j$. A combination of the
fusion hierarchy and truncation identity allows one to obtain an equation
for $\tau^{(2)}(\lm)=t(\lm)$. This method was applied to many integrable
models, in particular, to the RSOS models in  \cite{BR89}
and to the root of unity  lattice vertex models
in  \cite{N2003}. The functional relations for the $\tau^{(2)}$-model for $N=3$
and the superintegrable  case  were first guessed in \cite{AMCP}
and have been solved to some extend in \cite{McCR90}.

The goal of the present section is to prove that the relations to determine the
values $E_1,\ldots,E_{n-1}$ entering $t(\lm)$ also have the form of a truncation identity.
We formulate this statement as follows:
\begin{theorem}\label{th2}
The polynomial $\tau^{(N+1)}(\lm)$ satisfies the ``truncation'' identity
\be\label{reltau}
\tau^{(N+1)}(\lm)-
\om^{\rho}\,z(\lm)\, \tau^{(N-1)}(\om\lm)=
{\cal A}_n(\lm^N)+{\cal D}_n(\lm^N)
\ee
if and only if the system of homogeneous equations (\ref{BAX}) for all $k$
has a nontrivial solution.
\end{theorem}
Note, the classical polynomial $\;{\cal A}_n(\lm^N)+{\cal D}_n(\lm^N)\;$ corresponds
to $\;\alpha_q+\bar \alpha_q\;$ in \cite{B_tau}.

\medskip
\noindent {\it\bf Proof.}
Let $t(\lm)$ be a polynomial \r{tnh}, \r{iE0En} such that
the systems of homogeneous equations (\ref{BAX}) for all $k$
have a nontrivial solution.
We shall show that the polynomial $P(\lm)=\tau^{(N+1)}(\lm)-
\om^{\rho}\,z(\lm)\, \tau^{(N-1)}(\om\lm)$ at the left-hand side of
(\ref{reltau}) is equal to ${\cal A}_n(\lm^N)+{\cal D}_n(\lm^N)$
With this aim we introduce the matrix
\[
\Gamma(\lm)\:=\:\lk\ba{cc}\tau^{(2)}(\lm)& \om^\rho z(\lm)\\ -1& 0\ea \rk.
\]
Then it is easy to verify from \r{rectau} by induction that
\[
\Gamma(\om^{j-1}\, \lm)\: \cdots\: \Gamma(\om \lm)\: \Gamma(\lm)\:=\:\lk\ba{cc}
\tau^{(j+1)}(\lm)& \om^\rho z(\lm)\, \tau^{(j)}(\om \lm)\\
-\tau^{(j)}(\lm)\;& -\om^\rho z(\lm)\, \tau^{(j-1)}(\om \lm)\ea \rk
\]
and we see that
\be\label{Gprod}
P(\lm)\;=\;\tr\: \Gamma(\om^{N-1}\, \lm)\: \cdots\: \Gamma(\om \lm)\: \Gamma(\lm)\,.\ee
This relation shows the invariance of $P(\lm)$ under cyclic shifting $\lm\to\om\lm$.
It means that in fact $P(\lm)$ depends only on $\lm^N$. We denote ${\cal P}(\lm^N)=P(\lm)$.
Thus we have to show that $\;{\cal P}(\lm^N)={\cal A}_n(\lm^N)+{\cal D}_n(\lm^N)$.
The direct calculation gives that the coefficients of $\lm^0$ and $\lm^{Nn}$
at both sides of this equation coincide.
In order to calculate the coefficient in front of $\lm^0$ in the trace
of the  product of $\Gamma$-matrices (\ref{Gprod}) one has to substitute
\be\label{Gzero}
\Gamma(\lm)\to \lk\ba{cc}1 +\om^\rho \prod_{m=1}^n b_m d_m/\kp_m \hq&
\om^\rho \prod_{m=1}^n b_m d_m/\kp_m\\ -1& 0\ea \rk
\ee
where only the lowest terms in $\lm$ in the matrix elements were kept.
Therefore the lowest term in $\lm$ in $P(\lm)$ is
\[
\tr \lk\ba{cc}1 +\om^\rho \prod_{m=1}^n \frac{b_m d_m}{\kp_m}&
\om^\rho \prod_{m=1}^n \frac{b_m d_m}{\kp_m}\\ -1& 0\ea \rk^N.
\]
Finding the eigenvalues of the matrix (\ref{Gzero}) one can easily calculate
the latter trace which is the lowest term ${\cal P}(0)$ and identify it
with the lowest term
\[
{\cal A}_n(0)+{\cal D}_n(0)=1+\prod_{m=1}^n \frac{b_m^N d_m^N}{\kp_m^N}
\]
of the polynomial ${\cal A}_n(\lm^N)+{\cal D}_n(\lm^N)$ calculated by means of relations
(\ref{Lclass}) and  (\ref{Tclass}).
The case of the coefficients in front of $\lm^{Nn}$ can be treated analogously.

Therefore to prove the Theorem~\ref{th2} it remains to prove
\be\label{Plmnk}
P(\lm_{n,k})={\cal A}_n(\lm_{n,k}^N)+{\cal D}_n(\lm_{n,k}^N), \qquad
k=1, 2,\ldots, n-1\,,
\ee
where $\lm_{n,k}$ are given by \r{roots} and
 $\lm_{n,k}^N=\epsilon r^N_{n,k}$ are zeros of the polynomial
(\ref{Brec}).
Let us fix some $k$ and $\rho_{n,k}$ and denote the matrix of the coefficients
of (\ref{BAX}) with respect to the variables
$\;\tilde q_k(\rho_{n,k})$, $\;\tilde q_k(\rho_{n,k}-1),\: \ldots$,
$\:\tilde q_k(\rho_{n,k}-N+1)\;$ by $\;\:{\cal M}$:\\ [-3mm]
\be\label{matM} {\cal M}\:=\:\lk\BAR{cccccc} \!t_0\;&-\Delta_1^-&0&\ldots&0&-\Delta^+_0\\
  -\Delta^+_1&\!t_1\;&-\Delta^-_2&\ldots&0&0\\  0&-\Delta^+_2&t_2&\ldots&0&0\\
 \multicolumn{6}{c}{\ldots \qquad \ldots\qquad \ldots}\\
 -\Delta^-_0&0&0&\ldots&-\Delta^+_{N-1}&t_{N-1}\! \EAR\rk, \ee
where we abbreviated:
$\;t_j=t(\om^j\lm_{n,k});\;\;\Delta^\pm_j=\Delta_\pm(\om^j\lm_{n,k})\,.$
In order (\ref{BAX}) to have a nontrivial solution,
the matrix $\cal M$ must be degenerate.
Let $\cal M'$ be the matrix which has the same matrix elements as $\cal M$
except for ${\cal M}'_{1,N}=0$ and ${\cal M}'_{N,1}=0$.
Then, using the recursive definition (\ref{rectau})
of $\tau^{(j)}(\lm)$ and \r{zDelta}, it is easy to show
that the principal minor corresponding to the first $j$, $j\le N$ rows
of the matrix $\:{\cal M'}\:$ gives $\:\tau^{(j+1)}(\lm_{n,k})$.
Calculating the determinant of the matrix \r{matM} and equating it to zero
we obtain
\bea \det {\cal M}&=&\tau^{(N+1)}(\lm_{n,k})-
\om^{\rho}z(\lm_{n,k})\tau^{(N-1)}(\om\lm_{n,k})\ny\\
&-&\!\!\av \Delta_+(\lm_{n,k} \om^s)
-\av \Delta_-(\lm_{n,k} \om^s)\,=\,0\,. \label{detM}\eea
Further,
\[\fl\av \Delta_-(\lm_{n,k} \om^s)=\chi_k^N (-1)^{n-1} r_{n,k}^{N(n-1)}\:
       \det {\cal L}_n(\lm_{n,k}^N)\;=\;
       \epsilon\frac{{\cal B}_{n-1}(\lm_{n,k}^N)}{r_n^N \lm_{n,k}^N}\;
       \det{\cal L}_n(\lm_{n,k}^N), \]
where we have used \r{Dpm}, \r{detlN},  \r{chi}, \r{xxx}, \r{Brec} and
\[ \chi_k^N\,=\,\frac{(-1)^n\: r_{n-1,0}^N}{r_n^N\; (r_{n,k}^N)^{n-1}}\;\:
\prod_{s=1}^{n-2}\;(r_{n-1,s}^N-r_{n,k}^N)\:=\:
\frac{(-1)^{n-1} {\cal B}_{n-1}(\lm_{n,k}^N)}
{r_n^N\: (r_{n,k}^N)^{n}}\,.
\]
Evaluating (\ref{calAB}) at $m=n$ and $\,\lm=\lm_{n,k}\,$ so that
$\,{\cal B}_{n}(\lm_{n,k}^N)=0\,,$ finally we obtain
\[
\av \Delta_-(\lm_{n,k} \om^s)={\cal A}_n(\lm_{n,k}^N).
\]
Substituting $\lm=\lm_{n,k}$ into
\bea \fl \quad\det\,{\cal T}_n(\lm^N)={\cal A}_n(\lm^N)\,{\cal D}_n(\lm^N)-
{\cal B}_n(\lm^N)\,{\cal C}_n(\lm^N) =
  \prod_{m=1}^n \det {\cal L}_m(\lm^N)\ny\\[-3mm]\fl\qquad\qquad
   = \prod_{m=1}^n \av F_m(\lm\,\om^{s-1}) = \av z(\lm\,\om^s) =\av \lk\,\Delta_+(\lm\,\om^s)
         \cdot\Delta_-(\lm\,\om^s)\,\rk.\eea
we get\\[-11mm]
\[
\av \Delta_+(\lm_{n,k} \om^s)={\cal D}_n(\lm_{n,k}^N).
\]
Using (\ref{detM}) we obtain (\ref{Plmnk}).

Conversely, if we have the polynomials $\tau^{(N-1)}(\lm)$ and $\tau^{(N+1)}(\lm)$
built from $\tau^{(2)}(\lm)=t(\lm)$ (see \r{tnh}, \r{iE0En}) by the recursion \r{rectau}
and satisfying \r{reltau} we get \r{detM} at particular values of $\lm$.
This means that the Baxter equations \r{BAX} have nontrivial solutions.

This completes the proof of Theorem~\ref{th2}.

\section{Periodic homogeneous \BBS model for $N=2$}
\label{perN2}
\subsection{Solution of the Baxter equations}

In this section we consider in more detail
 the case of the $N=2$ periodic homogeneous \BBS model, where $\;\om=-1$.
By homogenous we mean that the parameters $a$, $b$, $c$, $d$ and $\kp$ are taken to be the same
 for all sites. As it was shown in \cite{Bugrij}, for
$N=2$ and with arbitrary homogeneous parameters this model is a particular case (``free
fermion point'') of the generalized Ising model.

We will find the eigenvalues $t_n(\lm|\rho,{\bf E})$ of the transfer-matrix
${\bf t}_n(\lm)$ using a functional relation (see also \cite{CPMBaxSol}, where
a similar calculation is presented). We use the short notation $\,t(\lm)\,$ for
$\:t_n(\lm|\rho,{\bf E})$.
From the previous section we have
\be\label{tlm}\fl
t(\lm)=1+(-1)^\rho\frac{b^n d^n}{\kp^n}+E_1\lm+\cdots+E_{n-1}\lm^{n-1}
+\lm^{n} (a^n c^n+(-1)^\rho \kp^n).
\ee
Using \r{rectau} for $j=2$ and eliminating $\tau^{(3)}$ by \r{reltau}
we get the functional relation
\be\label{frN2}
t(\lm)\:t(-\lm)= (-1)^\rho (z(\lm)+z(-\lm))+{\cal A}_n(\lm^2)+{\cal D}_n(\lm^2)
\ee  which we shall use to find $t(\lm)$. Equivalently, we could have obtained
\r{frN2} by multiplying together the two
Baxter equations \r{BAX} for $\lm_{n,k}=\pm r_{n,k}$.

Postponing for a moment the derivation (which will be supplied after \r{acac}),
let us anticipate that \r{frN2} can be rewritten as
\be\label{ttABC}
t(\lm)\:t(-\lm)=(-1)^n \prod_q
\left ( A(q)\,\lm^2\,-\,C(q)\,+2{\rm i}\,B(q) \lm\right),
\ee
where\\ [-10mm]
\bea  \lefteqn{ A(q)\,=\,a^2\,c^2\,-2\kp\, ac\,\cos q\,+\kp^2;\hs
B(q)\,=\,(ad-bc)\sin q\,;}\ny\\ [2mm] &&
\hs\hs C(q)=1-2\frac{bd}{\kp}\cos q + \frac{b^2d^2}{\kp^2}\,,\eea
$q$ is running over the set $\;\pi (2s+1-\rho)/n\,$, $\,s=0,1,\ldots,n-1$.
Factorizing (\ref{ttABC}) we get
\be t(\lm)\,t(-\lm)=(-1)^n\prod_q A(q)(\la\,-\la_q)(\la\,+\la_{-q}) \label{abc} \ee
with   \\ [-10mm]
\be  \lm_q=\frac{1}{A(q)}(\sqrt{D(q)}\,-{\rm i} B(q)),\qquad
D(q)\;=\;A(q)\,C(q)\,-B(q)^2.        \label{acb}\ee
We fix the sign of $\:\sqrt{D(q)}\:$ by the conditions
\bea  \sqrt{D(q)}&=&\sqrt{D(-q)},\qquad q\ne 0,\pi;
\label{dq1}\\ [2mm]
 \sqrt{D(0)}&=&(\kp\,-ac)(1\,-bd/\kp);\quad \sqrt{D(\pi)}\;=\;(\kp\,+ac)(1\,+bd/\kp).
\label{dq}\eea
In what follows we shall need the relations
\be\label{prodA} \prod_{q} A(q)\,=\, \prod_{q}
(\kp-e^{{\rm i}q}ac)(\kp-e^{-{\rm i}q}ac)= (a^n c^n+(-1)^\rho\kp^n)^2\,,\ee
\be\label{eqDB}
\prod_q (\sqrt{D(q)}-{\rm i} B(q)) =
\prod_q (\kp-e^{{\rm i}q}ac)(1-e^{{\rm i}q} bd/\kp).
\ee
The last relation can be obtained by grouping terms with
opposite signs of $q$ (modulo $2\pi$) and using the definition of $\sqrt{D(q)}$.
Using (\ref{prodA}) we get
\bea
t(\lm)t(-\lm)&=&(-1)^n (a^n c^n+(-1)^\rho\kp^n)^2
\prod_{q} (\lm-\lm_q)\cdot \prod_{q}(\lm+\lm_{-q})\ny\\
&=&(-1)^n (a^n c^n+(-1)^\rho\kp^n)^2 \prod_{q} (\lm^2-\lm_q^2),
\label{ttlm}
\eea
where we made the change $q\to -q$ in second product.
From (\ref{tlm}) it follows that $t(\lm)$ can be presented as
\[
t(\lm)=(a^n c^n+(-1)^\rho\kp^n)\prod_{s=1}^{n}(\lm-\Lambda_s)
\]
with zeroes $\Lambda_s$. Therefore
\[
t(\lm)\:t(-\lm)=(-1)^n(a^n c^n+(-1)^\rho\kp^n)^2\:\prod_{s=1}^{n}\:(\lm^2-\Lambda_s^2).
\]
Comparing with (\ref{ttlm}) we obtain
\be\label{tNSR}
t(\lm)=(a^n c^n+(-1)^\rho\kp^n)\prod_{q} (\lm\pm\lm_q),
\ee
where the signs are not yet fixed. To fix these signs we compare
the $\la$-independent term in (\ref{tlm})
with the corresponding term in (\ref{tNSR}). The latter can be found using
\bea  \fl \lefteqn{(a^n c^n+(-1)^\rho\kp^n) \prod_{q} \lm_q\;=\;
(a^n c^n+(-1)^\rho\kp^n)^{-1} \prod_{q} (\sqrt{D(q)}-{\rm i} B(q))}\ny\\
&&\hspace*{-15mm}=\;(a^n c^n+(-1)^\rho\kp^n)^{-1}
\prod_{q} (\kp-e^{{\rm i}q}ac)(1-e^{{\rm i}q} bd/\kp)\;=\;
(-1)^\rho+ b^nd^n/\kp^n\,,\label{acac}\eea
where we took into account (\ref{eqDB}).
Therefore the number of minus signs in (\ref{tNSR}) must be even for the sector
$\rho=0$ and odd for $\rho=1$.
Thus we have obtained $2^n$ eigenvalues
($2^{n-1}$ each for both $\rho=0$ and $\rho=1$).
These eigenvalues provide the existence of nontrivial solutions of the system
\r{BAX} of homogeneous equations. These solutions give
the eigenvectors (\ref{EV}), a basis in the space of states of the
periodic \BBS model for $N=2$.

We conclude this section supplying the derivation of (\ref{ttABC}) from (\ref{frN2}):
Using
\[\fl
\delta_+(\lm):={(b+a \kp \lm)(d- c\kp \lm)}/{\kp},\qquad
\delta_-(\lm):=\delta_+(-\lm)={(b-a \kp \lm)(d+ c\kp \lm)}/{\kp}
\]
we easily verify the relations
$\delta_+^n(\lm)=z(\lm)$, $\delta_-^n(\lm)=z(-\lm)$,
$\delta_+(\lm)\delta_-(\lm)= \delta(\lm^2)$,
where $z(\lm)$ and $\delta(\lm^2)$ are given by (\ref{zDelta}) and
(\ref{deltalm}) respectively.
Taking into account
${\cal A}_n(\lm^2)+{\cal D}_n(\lm^2)=\tr \left({\cal L}(\lm^2)\right)^n=
x_+^n(\lm^2)+x_-^n(\lm^2)$ and the relations
\[\fl
\tau(\lm^2)= \tr{\cal L}(\lm^2)= x_+(\lm^2)+ x_-(\lm^2),\qquad
\delta(\lm^2)= \det{\cal L}(\lm^2)= x_+(\lm^2) x_-(\lm^2),
\]
we can rewrite the functional relation  (\ref{frN2}) as
\bea
\fl \qquad\quad t(\lm)t(-\lm)&=& (-1)^\rho (z(\lm)+z(-\lm))+{\cal A}_n(\lm^2)+{\cal D}_n(\lm^2)\ny\\
&=& (-1)^\rho \left(\delta_+^n+\delta_-^n\right)+x_+^n+x_-^n \ny\\ [3mm]
&=&(-1)^\rho\left( x_+^n+(-1)^\rho\delta_+^n\right)
\left((x_-/\delta_+)^n+ (-1)^\rho\right)\ny\\ [4mm]
&=&(-1)^\rho\prod_{q}\; (x_+ -e^{{\rm i}q} \delta_+)(x_-/\delta_+-e^{{\rm i}q})\ny\\
&=&(-1)^{n} \prod_{q}\; (\: e^{{\rm i}q} \delta_+-\tau(\lm^2)+
e^{-{\rm i}q} \delta_-\:)\ny\\ [-2mm]
&=&(-1)^n \prod_{q}\;
\lb\lk(a^2c^2+\kp^2)\lm^2-\frac{b^2d^2}{\kp^2}-1\rk
+2\lk\frac{bd}{\kp}-\lm^2\kp\, a\, c\rk\cos q \right.\ny\\ [-2mm]
 &&\hspace*{3cm}\left.-2{\rm i}\,\lm (a\,d\,-b\,c)\,\sin q\rb,
\eea
which confirms (\ref{ttABC}).

\subsection{Relation to the standard Ising model notations}

In the homogeneous $N=2$ case we have $\om=-1$ and $\hu_k^{-1}=\hu_k$,
so the cyclic $L$-operator \r{bazh_strog} reduces to
\be
L_k(\lm)=\lk\ba{cc}
1+\lm\, \kp\, \hv_k &  \lm \,\hu_k\, (a\,-b \hv_k)\\
\hu_k\, (c\,-d\, \hv_k) & \lm a c + \hv_k\, b\, d/\kp\ea \rk.\ee
Let us make the special choice of the parameters $\;d=bc/a\;$ and $\:\lm=b/(a\kp)$. Then
\be \fl L_k(\lm)=(1+{\bf v}_k\,b/a)\lk\ba{cc}
1 &  \hu_k {b}/{\kp}\\c \hu_k & {b c}/{\kp }\ea \rk\,=\,
(1+{\bf v}_k \,b/a)
\lk \ba{c} 1\\ c \hu_k \ea \rk
\lk \ba{cc} 1\; &  \hu_k {b}/\kp\ea \rk.\ee
 and the transfer-matrix is
\be\fl
{\bf t}_n(\lm)=\tr\; L_1(\lm)\,L_2(\lm)\cdots L_n(\lm)=
\prod_{k=1}^n (1+{\bf v}_k \cdot {b}/{a})\cdot
\prod_{k=1}^n (1+\hu_{k-1}\hu_{k}\cdot bc/\kp).
\ee
Recall that due to the periodic boundary conditions $\hu_{n+1}\equiv\hu_{k}$.
Using
\[\fl
\exp (K_1 \hu_{k-1}\hu_{k})=\cosh K_1 (1\!+\hu_{k-1}\hu_{k} \tanh K_1);\;
\exp (K^*_2 \hv_k)=\cosh K^*_2 (1\!+\hv_k \tanh K^*_2),
\]
and writing $\:\hu_k\,=\:\sigma^z_k\,$ and $\:\hv_k\,=\:\sigma^x_k$,
it is easy to identify ${\bf t}_n(\lm)$ with the standard Ising
transfer-matrix:
\[ \fl
{\bf t}_{\rm Ising}=\exp{\lk \sum_{k=1}^n\, K^*_2\, \sigma^x_k\rk}\:
\exp{ \lk\sum_{k=1}^n\, K_1 \,\sigma^z_{k-1}\,\sigma^z_k\rk};\quad\;
\tanh K^*_2=\frac{b}{a};\quad\; \tanh K_1=\frac{bc}{\kp}\,.
\]

\section{Conclusion}\label{discus}

This paper is devoted to the solution of the eigenvalue and eigenvector problems
for the finite-size inhomogenous periodic Baxter-Bazhanov-Stroganov
quantum chain model. We use an approach which had been developed in full detail for
the quantum Toda chain in \cite{KarLeb1,KarLeb2} and in \cite{KhLS} for
the relativistic deformation of the Toda chain. This approach consists of two
main steps: In order to find eigenvectors for the transfer matrix $A_n(\lm)+D_n(\lm)$
first we find the eigenvectors of the off-diagonal operator $B_n(\lm)$ adapting
the well known recurrent procedure described in \cite{KarLeb2}
to our root-of-unity case. Then, using these
eigenvectors, we construct the eigenvectors for the \BBS transfer matrix and show that the
coefficients of the decompositions of one set of eigenvectors in terms of the other set
factorizes into a product of single variable functions, each satisfying the Baxter
type equation. We show that the condition for these equations to have nontrivial solution
is equivalent to the functional relations for the transfer matrix eigenvalues in
the \BBS or $\tau^{(2)}$ model. In case of $N=2$ the Baxter equation can be solved
and as result we obtain the eigenstates of the transfer matrix of the generalized Ising model
at the  free fermion point. We shortly give the relation of the $N=2$ \BBS model parameters
to the standard Ising model parametrization.

\section*{Acknowledgements}
This work was partially supported by the grant
INTAS-OPEN-03-51-3350, the
Heisenberg-Landau program and France--Ukrainian project ``Dnipro'' and the grant
RFBR-05-01-086.

\appendix
\setcounter{section}{1}
\section*{Appendix: Amplitudes $\:r_{m,k}\:$ in the homogeneous case}

The determination of the amplitudes $r_{m,k}$ for the inhomogenous
\BBS model had been reduced to solving Equation \r{Brec} with \r{relBm}
of Section \ref{recinh}.
Here we show that  in the {\it homogenous} case this task simplifies to
solving just one quadratic equation, using a trigonometric parametrization.

In the homogeneous case we have
\be \fl a_m=a,\;\;b_m=b,\;\;c_m=c,\;\;d_m=d,\;\;\kp_m=\kp,\;\;r_m=r,\;\;
{\cal L}_m(\lm^N)={\cal L}(\lm^N)\label{hom}\ee and\\[-9mm]
\be\label{calABCDhom}
\left(\ba{cc}
{\cal A}_m(\lm^N) &  {\cal B}_m(\lm^N)\\
{\cal C}_m(\lm^N) &  {\cal D}_m(\lm^N)
\ea\right)\;=\;\left({\cal L}(\lm^N)\right)^m\!.
\ee
Using the fact that a $2\times 2$ matrix $\:\bf M\:$ with eigenvalues
$\mu_+$ and $\mu_-$ satisfies
\[
{\bf M}^m\;=\:\frac{\mu_+^m-\mu_-^m}{\mu_+-\mu_-}\;{\bf M}
\:-\,\frac{\mu_+^m \mu_--\mu_-^m \mu_+}{\mu_+-\mu_-}\;{\bf 1}\,,
\]
from the matrix $\;{\cal L}(\lm^N)\;$ we obtain
\bdm {\cal B}_m(\lm^N)\;=\;-\epsilon\, \lm^N r^N\,
\frac{x_+^m-x_-^m}{x_+-x_-},
\edm
where $x_+(\lm^N)$ and $x_-(\lm^N)$ are the eigenvalues of $\:{\cal L}(\lm^N)$.
These eigenvalues are the roots of the characteristic polynomial
$\;x^2\,-\tau(\lm^N)\, x\, +
\delta(\lm^N)\:=\,0\,$:
\be x_{\pm}\;=\;\frac{1}{2}(\,\tau\pm \sqrt{\tau^2\,-4\delta}\,),\label{xpm}  \ee
where, see \r{detlN},
\be\label{taulm}
\tau(\lm^N)=\tr {\cal L}(\lm^N)=
1+\frac{b^N d^N}{\kp^N} - \epsilon\lm^N (\kp^N + a^N c^N), \ee
\be\label{deltalm} \delta(\lm^N)= \det\, {\cal L}(\lm^N)\: =\:
(b^N/\kp^N-\epsilon\lm^N\,a^N)\,(d^N-\epsilon\lm^N\,c^N\,\kp^N). \ee

Introducing the variable $\phi$ by $x_+/x_-=e^{{\rm i}\phi}$ we
find that roots of ${\cal B}_m$ correspond to roots $\phi_{m,s}$
of $e^{{\rm i}m\phi}=1$ (without $\phi=0$):
\be
\label{zerphi} \phi_{m,s}=2\pi s/m, \qquad s=1,2,\ldots,m-1.
\ee
Now we need to find the explicit relation between $\lm^N$ and
$\:\phi$. We have
\be \tau+\sqrt{\tau^2-4\delta}\;=\;e^{{\rm i}\phi}\, (\tau-
\sqrt{\tau^2-4\delta})\hs\mbox{or}\hs
\tau^2=4 \delta \cos^2 \frac \phi 2\label{tdphi}.\ee
 Taking into account (\ref{taulm}) and (\ref{deltalm}), we
consider (\ref{tdphi}) as a quadratic equation for $\lm^N$:
\bea \fl \lefteqn{\lm^{2N} (a^{2N} c^{2N}+ \kp^{2N}- 2 a^N
c^N\kp^N\cos\phi)\;+(b^{2N} d^{2N}+ \kp^{2N}- 2 b^N
d^N\kp^N\cos\phi)/\kp^{2N}}\ny\\[2mm]
\fl &-& 2\epsilon \lm^N \lk(a^N-b^N)(c^N-d^N)+\frac{a^N b^N c^N
d^N}{\kp^N}+\kp^N - (a^N d^N+b^N c^N)\cos\phi \rk\;=\;0\,. \eea
The solution $\lm^N(\phi)$ of this equation describes the relation
between the variables $\lm^N$ and $\phi$. Therefore we can
translate the zeroes (\ref{zerphi}) of $\:{\cal B}_m(\lm^N(\phi))$
in terms of variable $\phi$, to zeroes $\,\lm^N(\phi_{m,s}).\:$ From
 (\ref{Brec}) we get
\be r_{m,s}^N=\epsilon \lm^N(\phi_{m,s}),\qquad s=1,2,\ldots,m-1. \label{ire0m} \ee
The value of $r_{m,0}^N$ can be found recursively from \r{rel0N} using \r{hom}:
\be \label{irel0N}
r_{m,0}^N= r_{m-1,0}^N\, a^N c^N+ r^N \kp^{N(m-1)}\,,\qquad
r_{1,0}^N= r^N=a^N-b^N\,.\ee

\section*{References}
\bibliographystyle{amsplain}

\end{document}